\documentclass{emulateapj}
\usepackage{amssymb}
\usepackage{apjfonts}
\usepackage{graphicx}
\usepackage{natbib}
\begin{document}
\title{Catalog of narrow $\rm C~IV$ absorption lines in BOSS (I): for quasars with $\rm z_{em} \leq 2.4$}
\shorttitle{Catalog of narrow $\rm C~IV$ absorption lines}
\shortauthors{Chen et al.}

\author{Zhi-Fu Chen\altaffilmark{1, 2}, Yi-Ping Qin\altaffilmark{1, 2, 3}, Cai-Juan Pan\altaffilmark{1}, Wei-Rong Huang\altaffilmark{2}, Ming Qin\altaffilmark{1}, Ha-Na Wu\altaffilmark{1}}

\altaffiltext{1}{Department of Physics and Telecommunication
Engineering of Baise University, Baise 533000 China;
zhichenfu@126.com}

\altaffiltext{2}{Center for Astrophysics,Guangzhou University,
Guangzhou 510006, China}

\altaffiltext{3}{Physics Department, Guangxi University, Nanning
530004, China}

\begin{abstract}
We have assembled absorption systems by visually identifying $\rm
C~IV\lambda\lambda1548,1551$ absorption doublets in the quasar
spectra of the Baryon Oscillation Spectroscopic Survey (BOSS) one by
one. This paper is the first of the series work. In this paper, we
concern quasars with relatively low redshifts and high
signal-to-noise ratios for their spectra, and hence we limit our
analysis on quasars with $z_{\rm em}\le2.4$ and on the doublets with
$W_r\lambda1548\ge0.2$ \AA.~ Out of the more than 87,000 quasars in
the Data Release 9, we limit our search to 10,121 quasars that have
the appropriate redshifts and spectra with high enough
signal-to-noise ratios to identify narrow C IV absorption lines.
Among them, 5,442 quasars are detected to have at least one $\rm
C~IV\lambda\lambda1548,1551$ absorption doublet. We obtain a catalog
containing 8,368 $\rm C~IV\lambda\lambda1548,1551$ absorption
systems, whose redshifts are within $z_{\rm abs}=1.4544$
--- $2.2805$. In this catalog, about 33.7\% absorbers have $0.2$
\AA$\le W_r\lambda1548<0.5$ \AA,~ about 45.9\% absorbers have $0.5$
\AA$\le W_r\lambda1548<1.0$ \AA,~ about 19.2\% absorbers have $1.0$
\AA$\le W_r\lambda1548<2.0$ \AA,~ and about 1.2\% absorbers have
$W_r\lambda1548\ge2.0$ \AA.
\end{abstract}

\section{Introduction}
Absorption lines are often observed in the quasar spectra, which are
a powerful tool to probe the gas in the Universe from high redshifts
to the present epoch (see Meiksin 2009 for a review). Quasar
absorption lines provide an unique chance to study the gaseous phase
(e.g., ionization states, kinematics, metallicities) of distant
galaxies that otherwise might be invisable, which are independent of
the luminosity of the background quasars. They are also important to
understand the star formation and evolution of the ordinary galaxies
(e.g., Prochter et al. 2006;  Zibetti et al. 2007; M\'enard et al.
2011; Chen 2013).

Narrow absorption lines (NALs), with the line width of a few hundred
$\rm km~s^{-1}$, can be classified into three categories according
to the relationship between the absorber and the corresponding
quasar. They are intrinsic absorption lines, associated absorption
lines and intervening absorption lines. The intrinsic absorption
lines are often believed to be physically related with the quasar
wind/outflow (e.g., Narayanan et al. 2004; Misawa et al. 2007;
Hamann et al. 2011). The associated absorption lines with $z_{\rm
abs} \approx z_{\rm em}$ probably arise from the gas in the quasar
host galaxy or the galaxy cluster around the quasar (e.g., Weymann
et al. 1979; Wild et al. 2008; Vanden Berk et al. 2008). The
intervening absorption lines with $z_{\rm abs} \ll z_{\rm em}$ are
due to the absorption of galaxies along the quasar sightlines
located at cosmological distances from the corresponding quasars
(e.g., Bahcall \& Spitzer 1969; Bergeron 1986; L\'opez \& Chen
2012). The criteria, determining whether the absorption lines are
truly tied to the corresponding quasars, is ambiguous, because there
are many factors that can disturb the observed absorption lines,
such as the signal to noise ratio of the quasar spectra. To day the
dividing line of the intervening absorption lines and the associated
absorption lines are usually derived by statistics (e.g., Richards
2001; Wild et al. 2008). The absorption lines at velocity
separations less than the value of $\rm \sim 0.02c - 0.04c$, when
compared to the quasar systems, are classified as associated
absorption line group (Vanden Berk et al. 2008; Wild et al. 2008).
However, that does not mean that narrow absorption lines with
velocity separation larger than that value completely belong to
intervening absorption lines. Narrow intrinsic absorption lines can
be formed in the quasar outflows with velocity separations up to,
and even exceeding $\rm 0.1c$ (e.g., Misawa et al. 2007; Tombesi et
al. 2011; Chen et al. 2013a; Chen \& Qin 2013).

$\rm C~IV\lambda\lambda1548,1551$ resonant doublets are observable
redward of the $\rm Ly\alpha1216$ emission line, which can be
detected over a redshift range of $z\approx1.5$ --- 5.5 in the
optical spectra. These lines are strong transitions and have good
profiles. They are valuable absorption lines to study the
intergalactic medium (e.g., Songaila \& Cowie 1996; Cowie \&
Songaila 1998; Songaila 2001; Schaye et al. 2003; Cooksey et al.
2010; D\'Odorico et al. 2010; Simcoe et al. 2011).

Based on the Sloan Digital Sky Survey (SDSS, York et al. 2000), many
previous works aimed at systematically searching for metal
absorption lines have been done (e.g., Quider et al. 2011; Qin et
al. 2013; Zhu \& M\'enard 2013; Cooksey et al. 2013). We are going
to identify absorption doublets, such as $\rm
C~IV\lambda\lambda1548,1551$ and $\rm Mg~II\lambda\lambda2796,2803$,
in the quasar spectra of the Baryon Oscillation Spectroscopic Survey
(BOSS), which is a part of the SDSS-III (Eisenstein et al. 2011). In
this paper, our work is to identify the $\rm
C~IV\lambda\lambda1548,1551$ absorption doublet, which becomes the
first in a series of papers on the absorption lines in the BOSS
quasar spectra.

In section 2, we show how we construct our $\rm
C~IV\lambda\lambda1548,1551$ absorption sample and present the
spectral analysis. The properties of the absorption lines are
presented in section 3. Section 4 is the discussion, and section 5
is the summary.

\section{Data analysis}
BOSS is the main dark time legacy survey of the third stage of the
SDSS (P\^aris et al. 2012; Eisenstein et al. 2011), which is a five
year programm. BOSS aims to get quasar spectra over $150,000$ with
$z_{\rm em}>2.15$ using the same $\rm 2.5 m$ telescope (Gunn et al.
2006; Ross et al. 2012) as the SDSS did. The spectra of BOSS span a
wavelength range of 3600 \AA --- 10400 \AA~ at a resolution of
$1300<R<3000$. The first data release of BOSS, SDSS Data Release
Nine (SDSS DR9), contains $87,822$ quasars detected over an area of
$3275~deg^2$ (P\^a aris et al. 2012).

In order to avoid the noisy region of the spectra, we exclude those
data shortward of 3800 \AA~ at the observed frame. The pair of $\rm
O~I\lambda1302$ and $\rm S~II\lambda1304$ has a wavelength
separation similar to that of the $\rm C~IV\lambda\lambda1548,1551$
doublet, and that may lead to misidentifications of the latter. To
avoid confusions arising from the $\rm Ly\alpha$ forest, and $\rm
O~I\lambda1302$ and $\rm S~II\lambda1304$ absorption lines, we
constrain our analysis on the wavelength range longward of 1310 \AA~
at the rest frame. We also conservatively constrain the upper
wavelength limit to $\rm 1548$\AA$\rm \times(1+z_{em})\times
\sqrt{(1-\beta)/(1+\beta)} $, where we adopt $\rm \beta=-1/30$ to
search for intervening $\rm C~IV\lambda\lambda1548,1551$ absorption
doublets. This cut reduces the quasar sample to $70,336$ quasars
with $z_{em}\gtrsim1.54$.

The noise superposed on the spectra with low signal-to-noise ratios
(SNR) often confuses the true absorptions. Here, we limit our
analysis to sources with high enough signal-to-noise ratios in the
surveyed spectral region. There is a median signal-to-noise
ratio (median SNR) of the spectrum of each quasar, which can roughly
reveal the level of the noise of the observation of the source.
Illustrated in Fig. 1 is the distribution of the median SNR of these
$70,336$ quasars. We find that the median value of this distribution
is quite close to $4$ (see Fig. 1). We accordingly adopt this value
to limit our analysis. That is, we select only quasars with $\rm
median~SNR\ge4$ in the surveyed spectral region.

\begin{figure}
\vspace{3ex} \centering
\includegraphics[width=7 cm,height=6 cm]{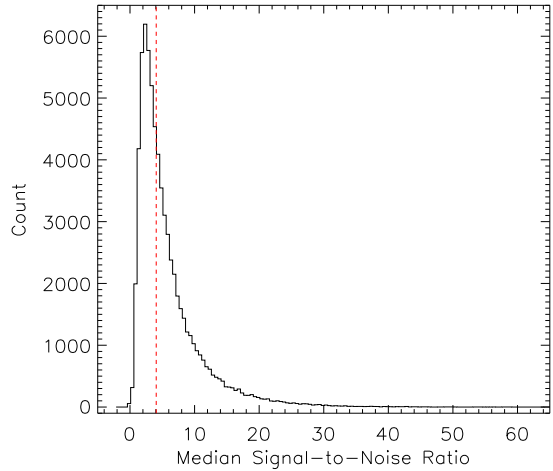}\vspace{3ex}
\caption{Distribution of the median SNR of the 70,336 quasars, in
the surveyed spectral region of $\rm C~IV\lambda\lambda1548,1551$
absorption doublets. The red dash line denotes the median value of
this distribution, which is located at 4.06.}
\end{figure}

As the first paper of the series of work, here we concern only
quasars with $z_{\rm em}\le2.4$. Taking into account all the above
limitations, we have 10,121 quasars with $1.54\lesssim z_{\rm
em}\le2.4$ to identify $\rm C~IV\lambda\lambda1548,1551$ absorption
doublets. The upper cuts of the emission redshift and the median SNR
are showed in Fig. 2. The distribution of emission redshifts of our
final quasar sample is plotted in Fig. 4.

\begin{figure}
\vspace{3ex} \centering
\includegraphics[width=7 cm,height=6 cm]{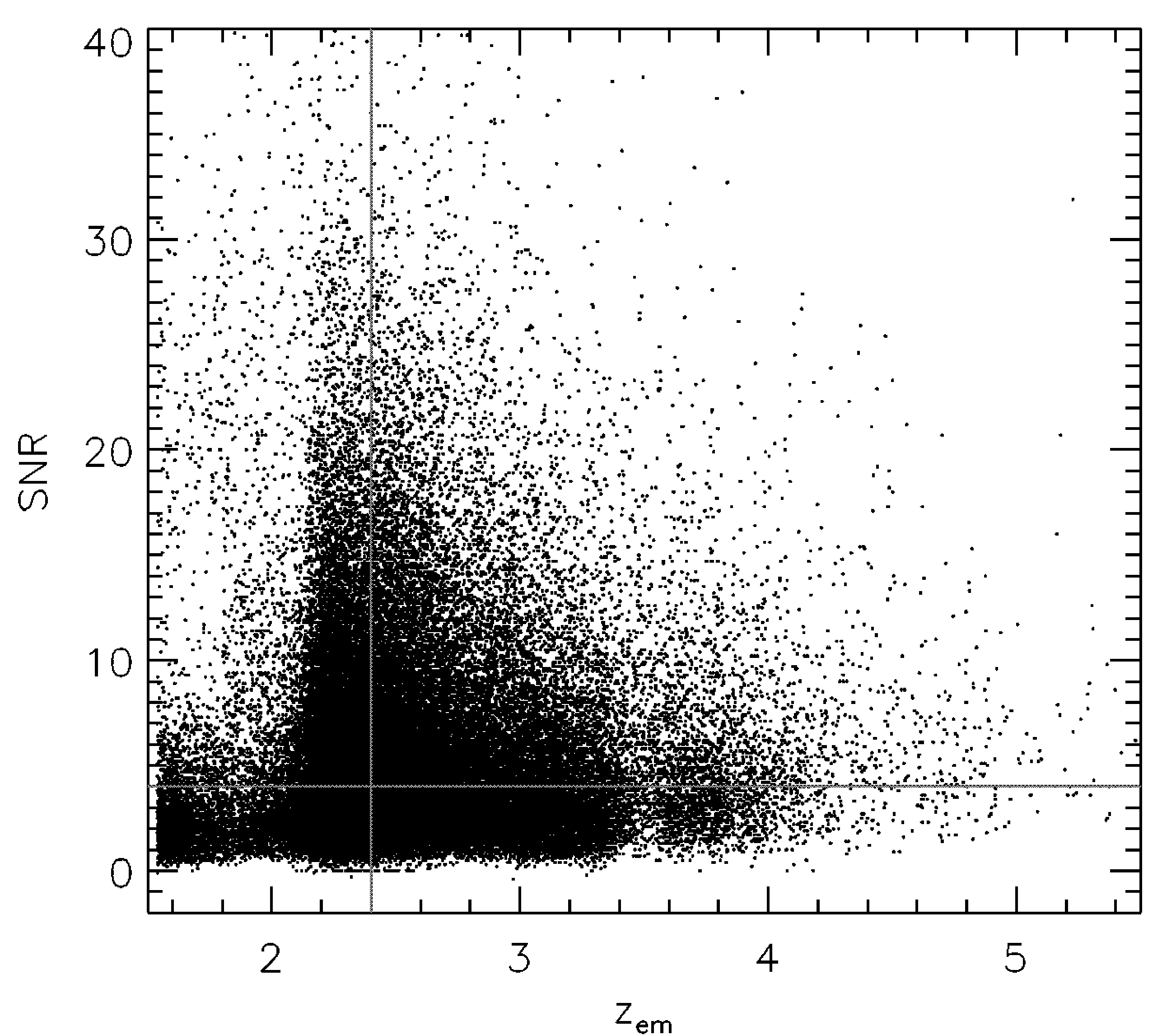}\vspace{3ex}
\caption{Plot of the median SNR, in the surveyed spectral region of
$\rm C~IV\lambda\lambda1548,1551$ absorption doublets, versus the
emission redshift of the $70,336$ quasars. The red lines are the
limits of SNR and $z_{em}$ used to construct our quasar sample.}
\end{figure}

We derive a pseudo-continuum for each quasar of our sample by
invoking a combination of cubic splines (for underlying continuum,
see Willian et al. 1992 for details) and Gaussians (for emission and
broad absorption features), which is utilized to normalize the
spectral data (fluxes and flux uncertainties). These processes are
iterated several times to improve the fittings of both the cubic
spline and Gaussian (e.g., Nestor et al. 2005; Quider et al. 2011,
Chen et al. 2013a,b). Shown in the left panels of Fig. 3 are several
quasar spectra (with various values of the median SNR) together with
their pseudo-continuum fitting curves. The pseudo-continuum
normalized spectra are presented in the right panels of Fig. 3.

We search $\rm C~IV\lambda\lambda1548,1551$ absorption candidates
from the pseudo-continuum normalized spectra. As the first step of
the searching (see also Chen et al. 2013a), the $2\sigma$ curve
below the pseudo-continuum fitting is marked, and then those
absorption figures located above this curve are ruled out.

In many cases, some very broad troughs appear in the blue wing of
$\rm C~IV$ or/and $\rm Si~IV$ emission lines. The broad absorption
line (BAL) is a confusing terminology. Based on the definition of
balnicity index (BI, Weymann et al. 1991), absorption troughs with
the width broader than $2000~\rm km~s^{-1}$ at depths $>10\%$ below
the pseudo-continuum fitting curve can be classified as BALs.
However, in terms of the absorption index (AI, Hall et al. 2002;
Trump et al. 2006), some narrower absorption troughs ($>1000~\rm
km~s^{-1}$) also belong to the BAL population. Knigge et al. (2008)
found that the BAL fraction will be underestimated in terms of BI,
and overestimated in terms of AI. They also found that both samples
of BI and AI show bimodal distributions, which bring about a problem
of the overlap of broad NALs and narrow BALs. We are going to
analyze only narrow absorption doublets with a few hundreds
$km~s^{-1}$, therefore, as the second step, we conservatively
disregard those absorption figures with widths broader than
$2000~\rm km~s^{-1}$ and at depths $>10\%$ below the
pseudo-continuum fitting curve in our program autonomically.

In the third step, each absorption trough is fitted by a Gaussian
component, and the absorption figures with the full width at half
maximum (FWHM) greater than $800~\rm km~s^{-1}$ are ruled out. And
then, we search the candidates of $\rm C~IV\lambda\lambda1548,1551$
absorption doublets from the residual absorption figures.

In the fourth step, we measure the equivalent widths ($W_r$) of
these candidate absorption lines at the rest-frame from the Gaussian
fittings, and estimate their uncertainties by
\begin{equation}
(1+z)\sigma_w=\frac{\sqrt{\sum_i
P^2(\lambda_i-\lambda_0)\sigma^2_{f_i}}}{\sum_i
P^2(\lambda_i-\lambda_0)}\Delta\lambda,
\end{equation}
where $P(\lambda_i-\lambda_0)$ is the line profile centered at
$\lambda_0$, $\lambda_i$ is the wavelength, and $\sigma_{f_i}$ is
the normalized flux uncertainty as a function of pixel (Nestor et
al. 2005; Chen et al. 2013b; Chen \& Qin 2013). The sum is performed
over an integer number of pixels that covers at least $\pm 3$
characteristic Gaussian widths. We adopt the method provided by Qin
et al. (2013) to evaluate the signal-to-noise ratio of the
absorption line for the candidates as well. $\rm 1\sigma$ noise is
calculated by:
\begin{equation}
\sigma_N=\sqrt{\frac{\sum\limits_{i=1}^M[\frac{F^i_{noise}}{F^i_{cont}}]^2}{M}},
\end{equation}
where $\rm F_{noise}$ is the flux uncertainty, $\rm F_{cont}$ is the
flux of the psuedo-continuum fit, and $i$ represents the pixel in
the wavelength range of
1548{\AA}$\rm\times(1+z_{abs})-5${\AA}$\rm<\lambda_{obs}<1551${\AA}$\rm\times(1+z_{abs})+5${\AA}.
The signal-to-noise ratio of the absorption line is determined by:
\begin{equation}
SNR^\lambda=\frac{1-S_{abs}}{\sigma_N},
\end{equation}
where $\rm S_{abs}$ is the smallest value of the normalized spectral
flux within an absorption trough. Finally, we select only the
absorption lines with $W_r>0.2$ \AA~ and $SNR^{\lambda} \ge 2.0$ for
both $\lambda1548$ and $\lambda1551$ lines. In this way, we get 8368
potential intervening $\rm C~IV\lambda\lambda1548,1551$ absorption
doublets. These absorption doublets are presented in Table 1.

\begin{figure*}
\centering
\includegraphics[width=7.6cm,height=1.5cm]{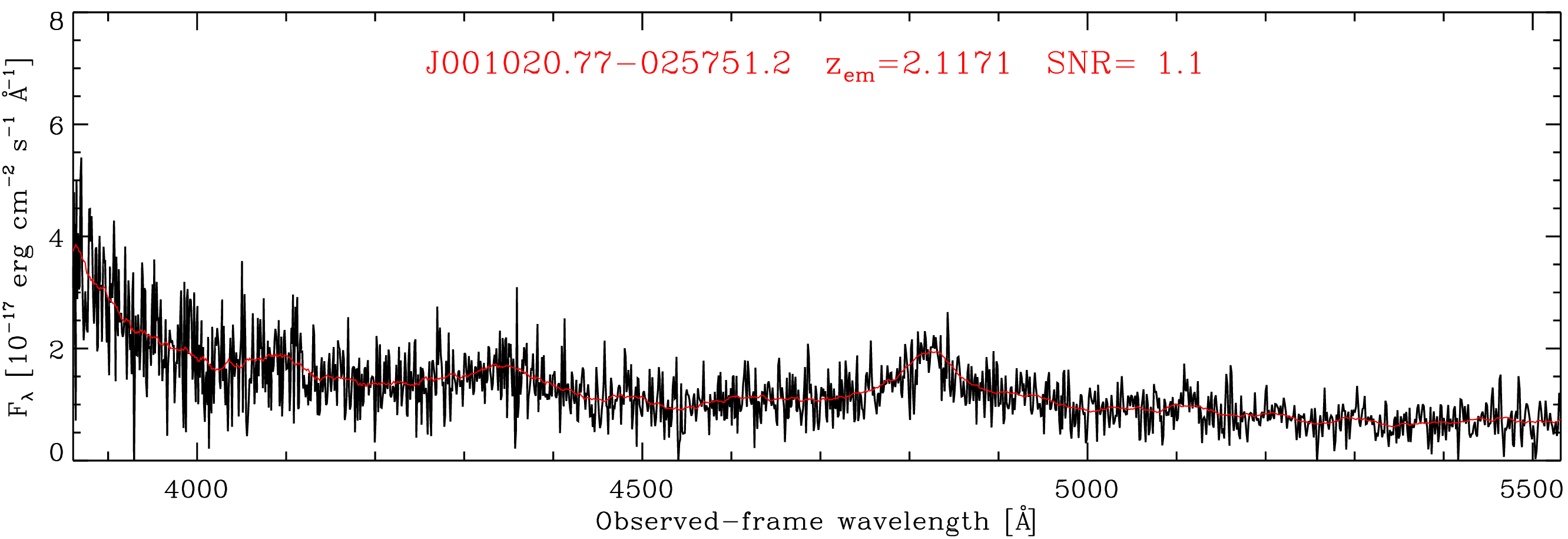}
\hspace{2ex}
\includegraphics[width=7.6cm,height=1.5cm]{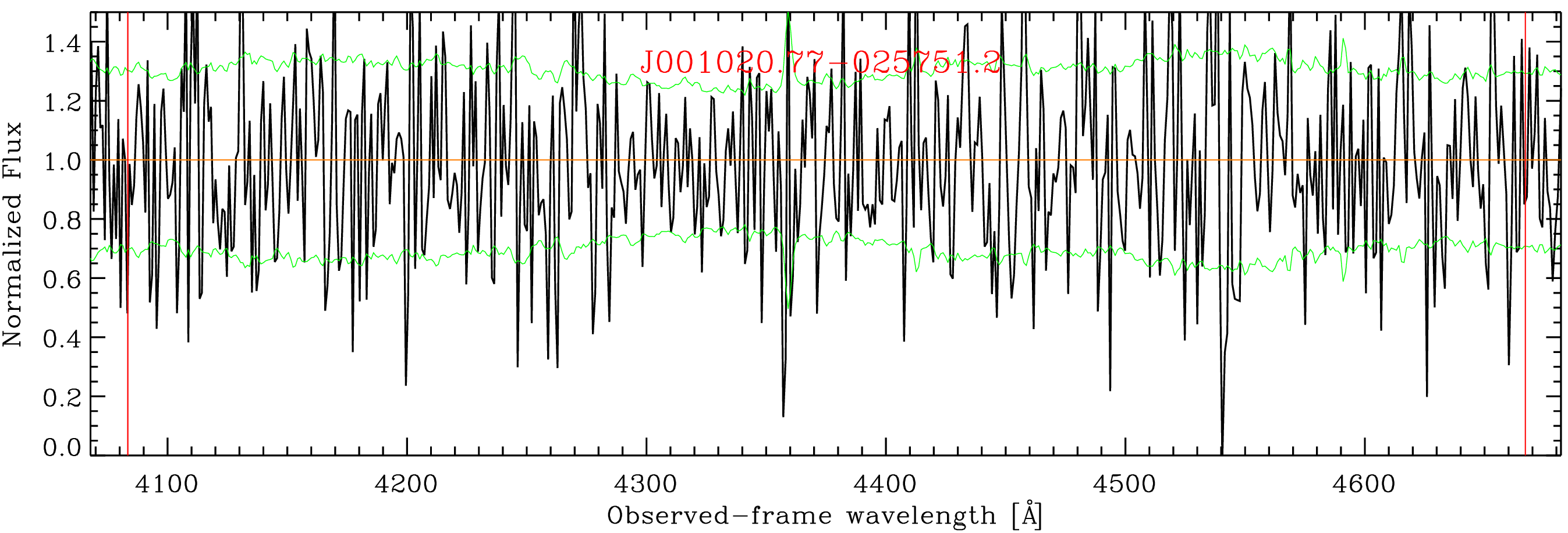}
\includegraphics[width=7.6cm,height=1.5cm]{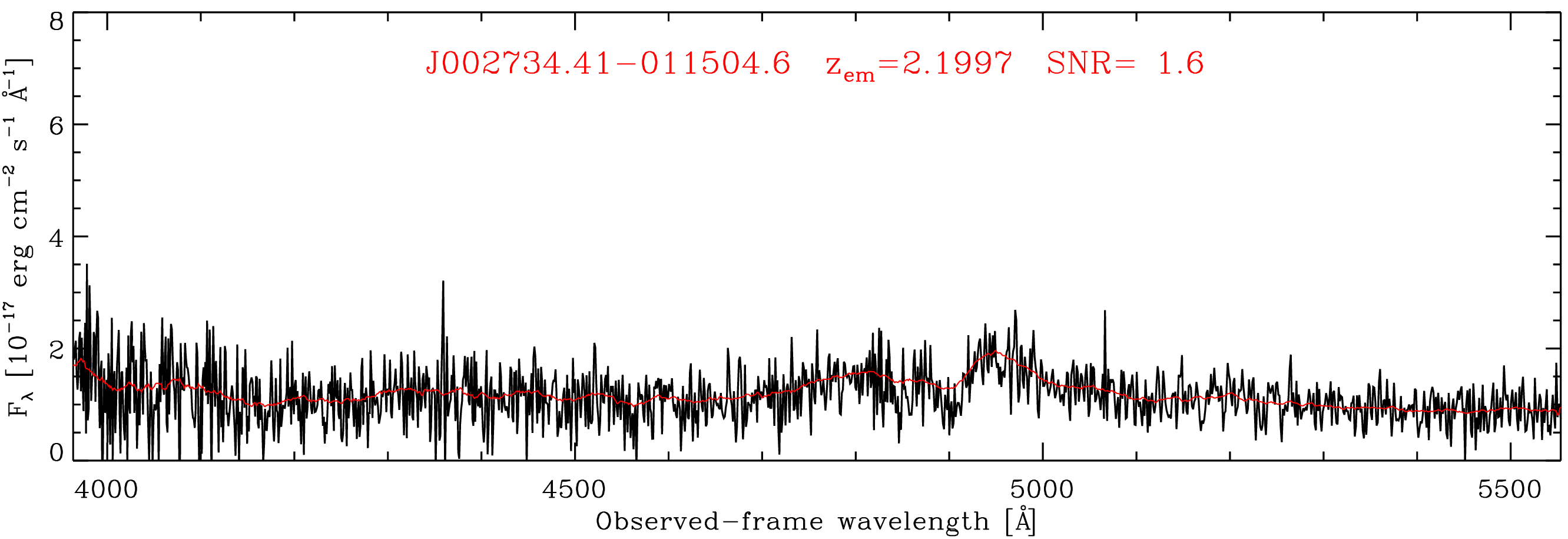}
\hspace{2ex}
\includegraphics[width=7.6cm,height=1.5cm]{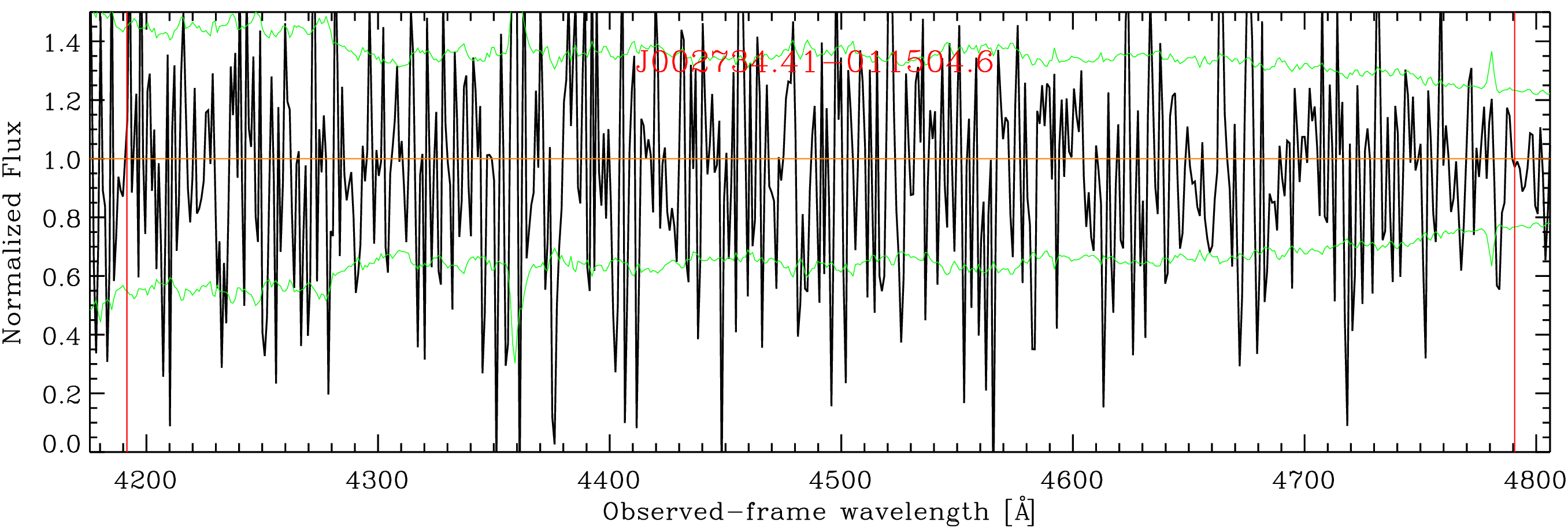}
\includegraphics[width=7.6cm,height=1.5cm]{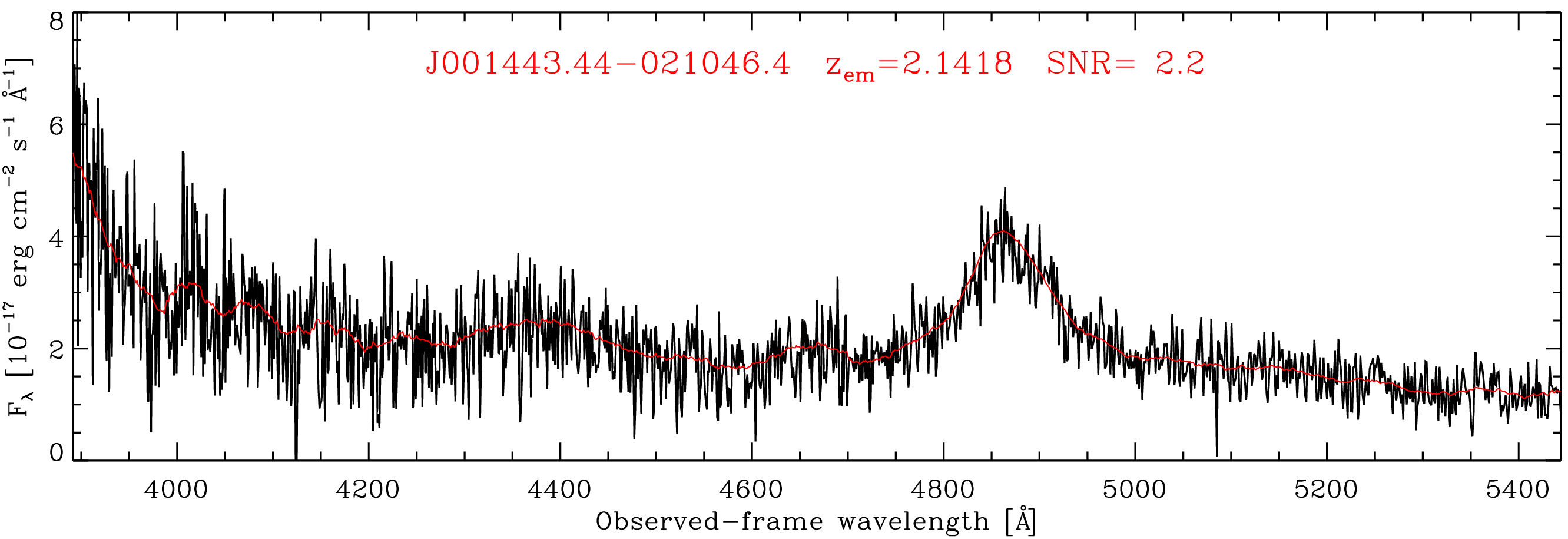}
\hspace{2ex}
\includegraphics[width=7.6cm,height=1.5cm]{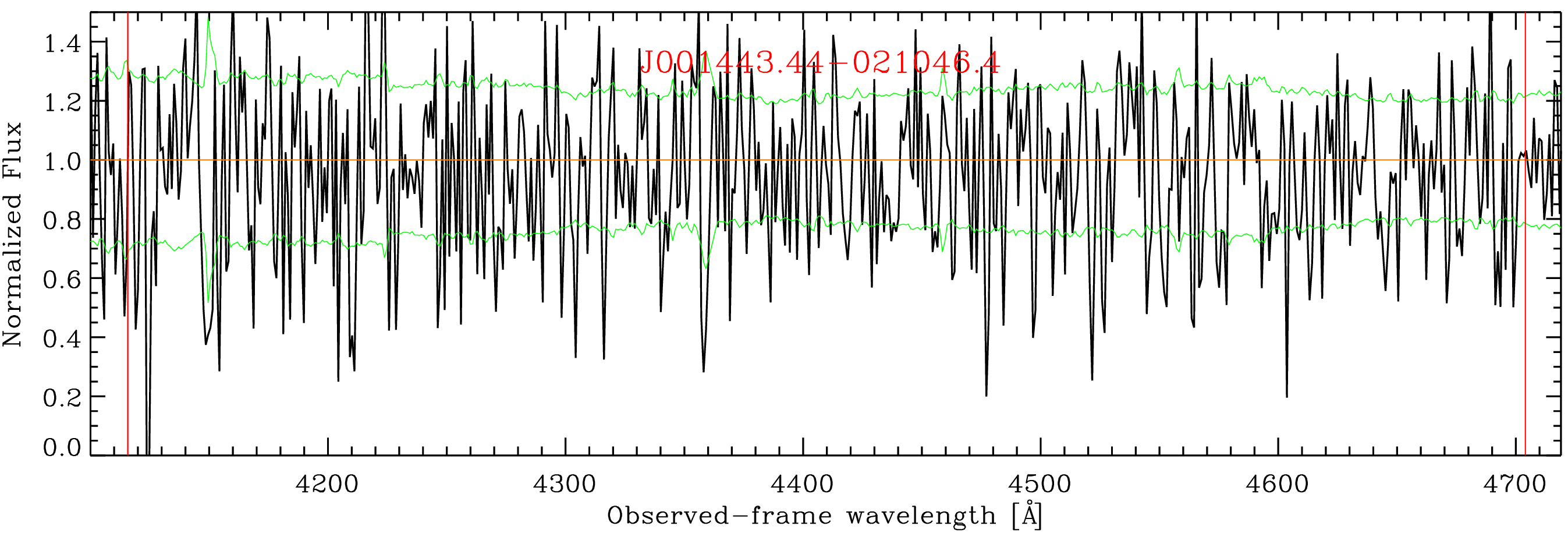}
\includegraphics[width=7.6cm,height=1.5cm]{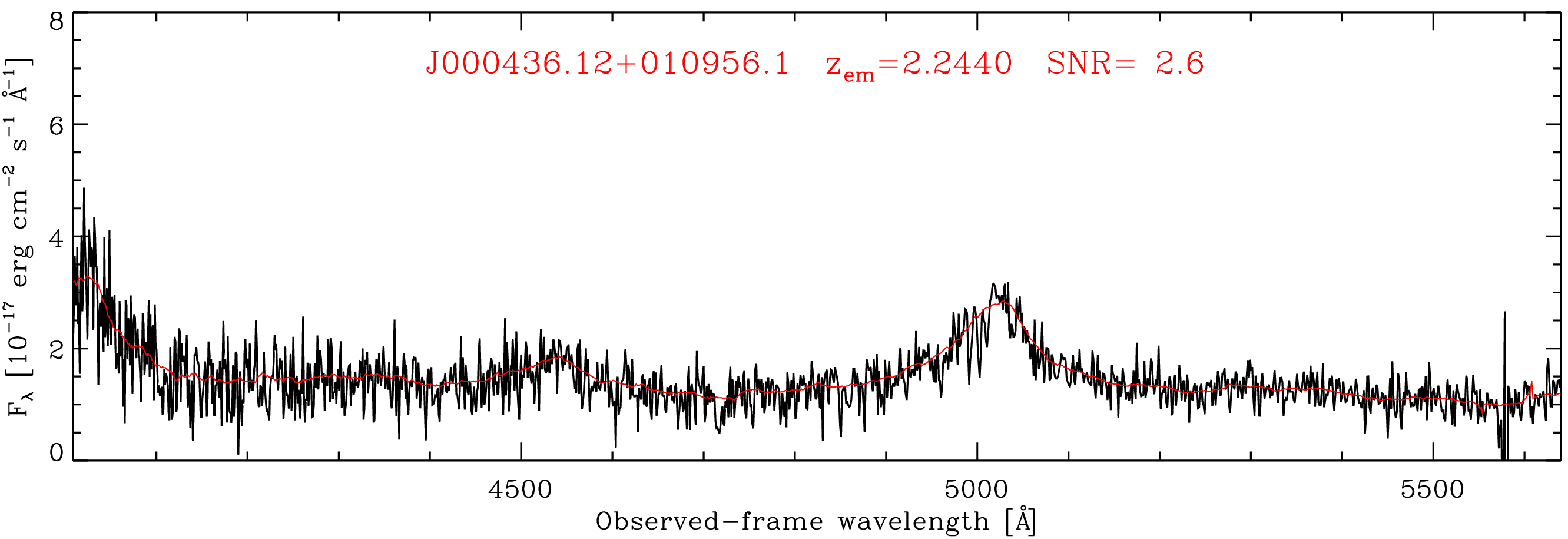}
\hspace{2ex}
\includegraphics[width=7.6cm,height=1.5cm]{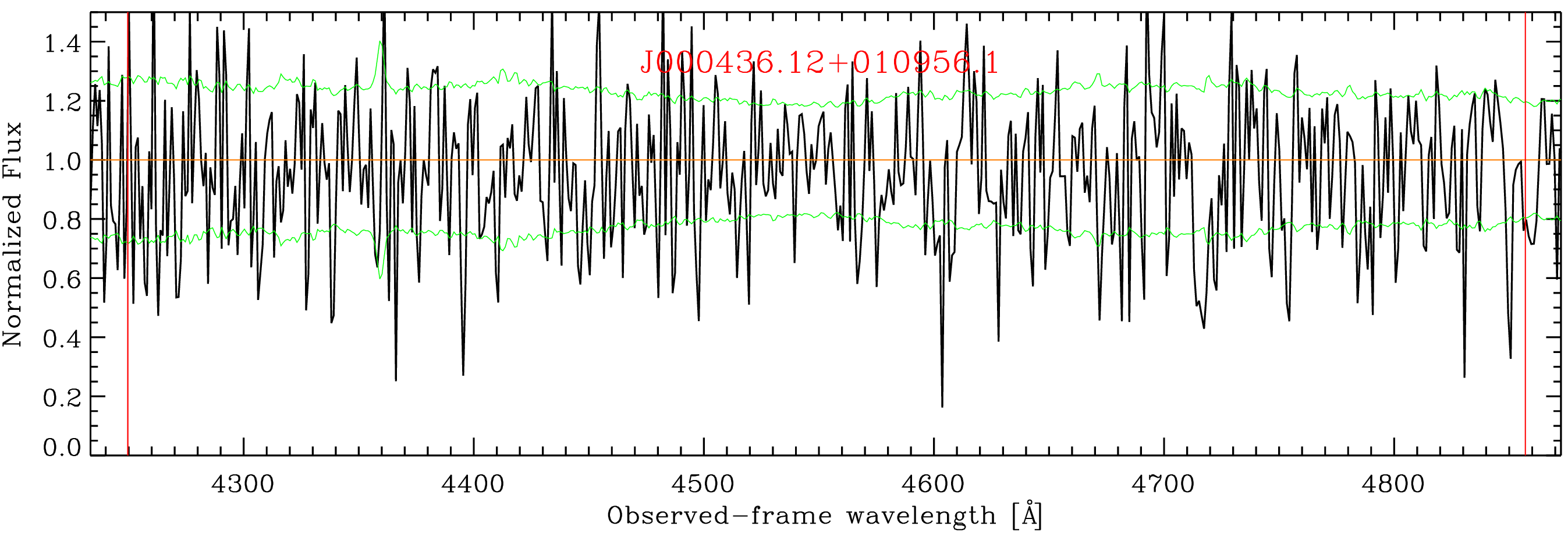}
\includegraphics[width=7.6cm,height=1.5cm]{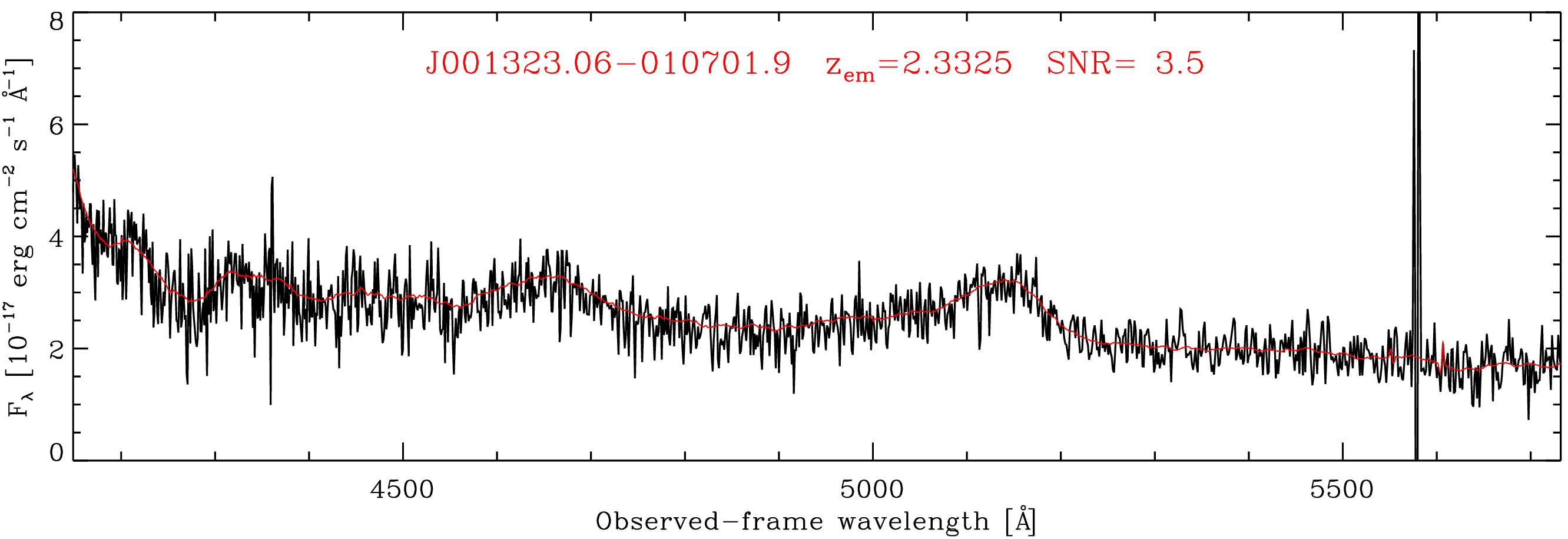}
\hspace{2ex}
\includegraphics[width=7.6cm,height=1.5cm]{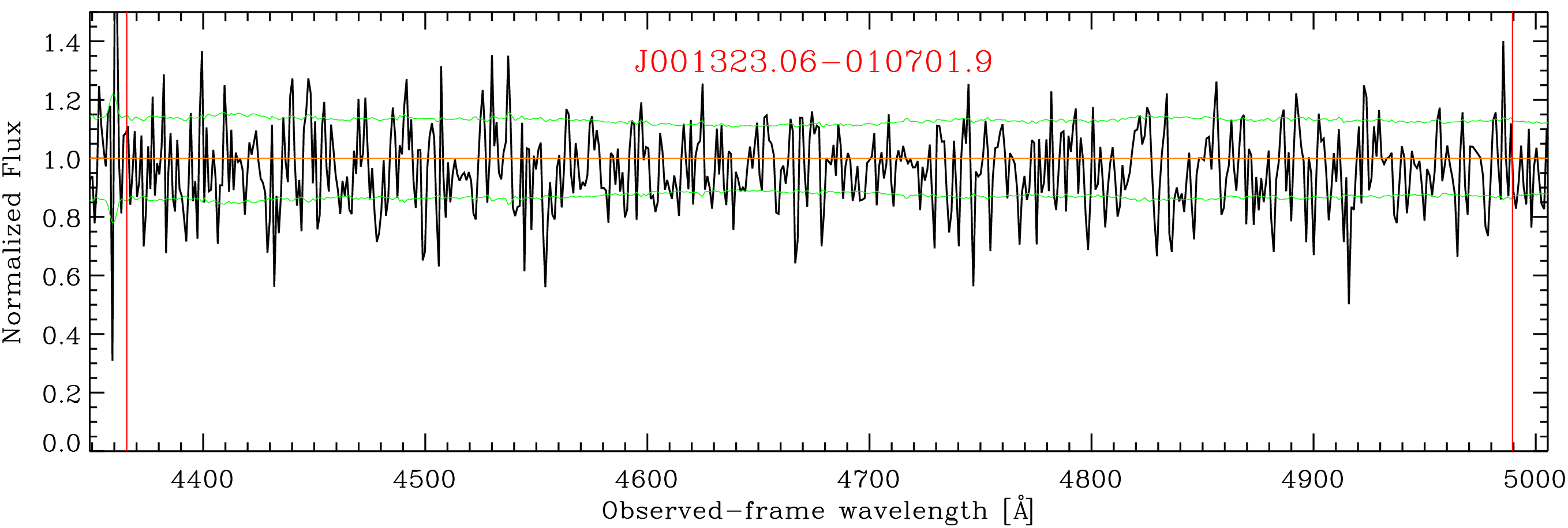}
\includegraphics[width=7.6cm,height=1.5cm]{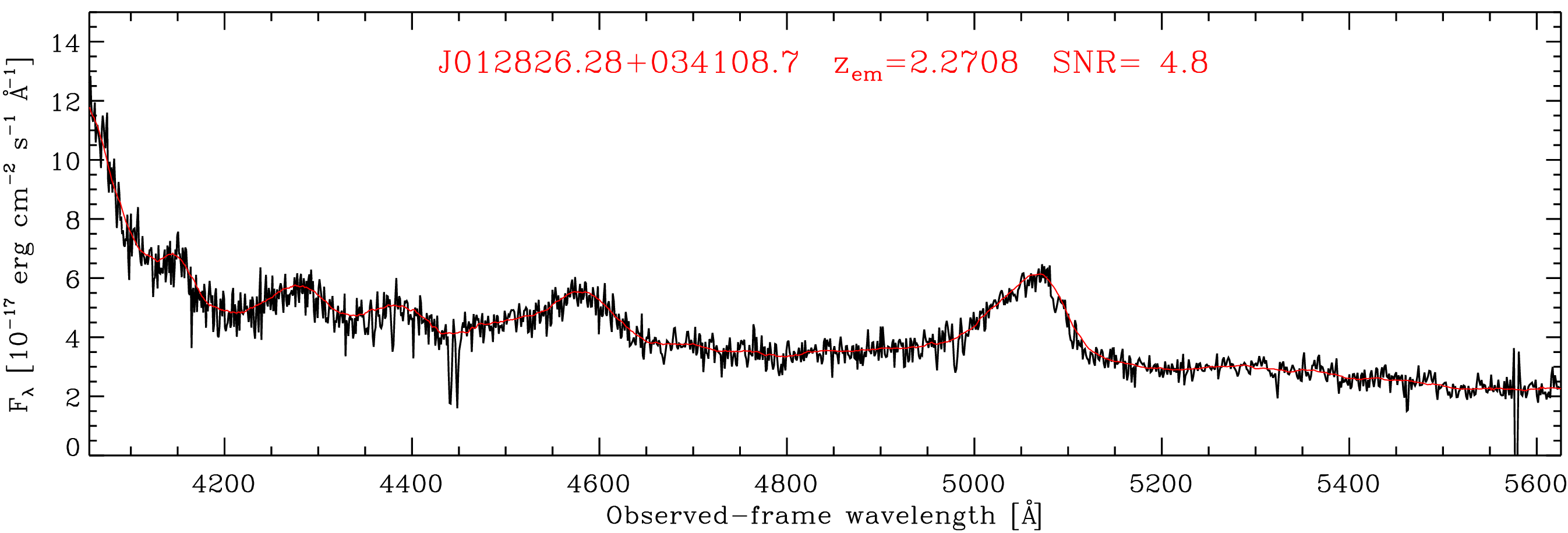}
\hspace{2ex}
\includegraphics[width=7.6cm,height=1.5cm]{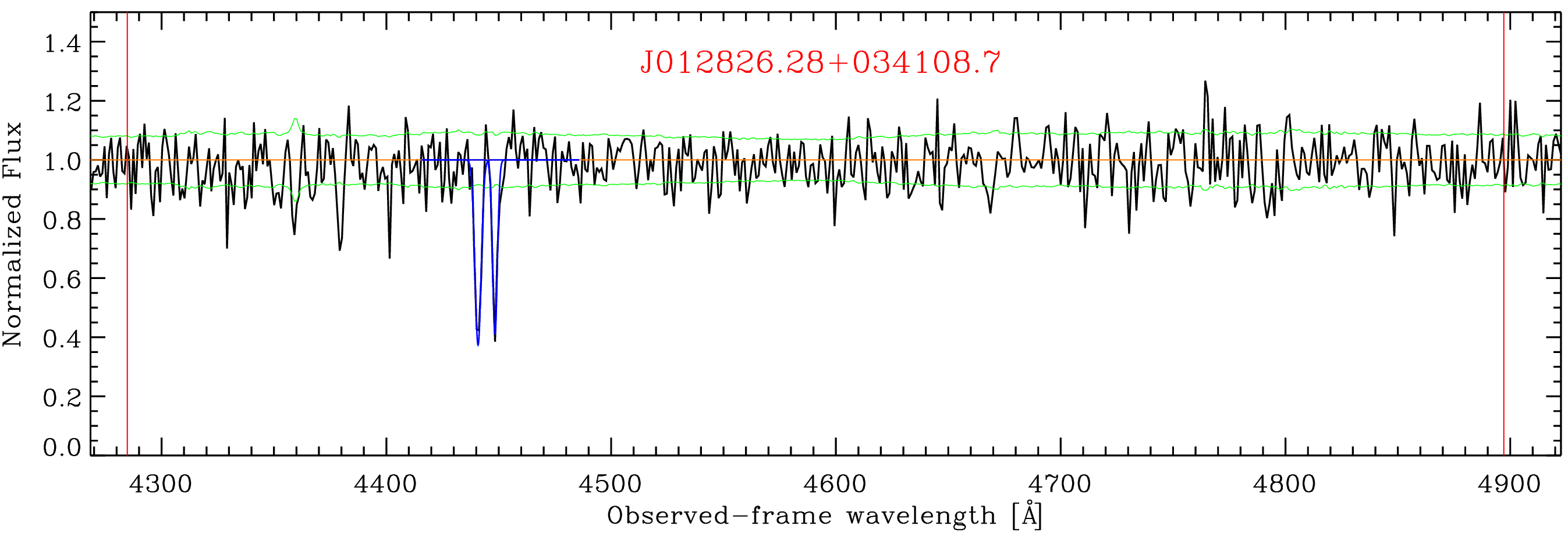}
\includegraphics[width=7.6cm,height=1.5cm]{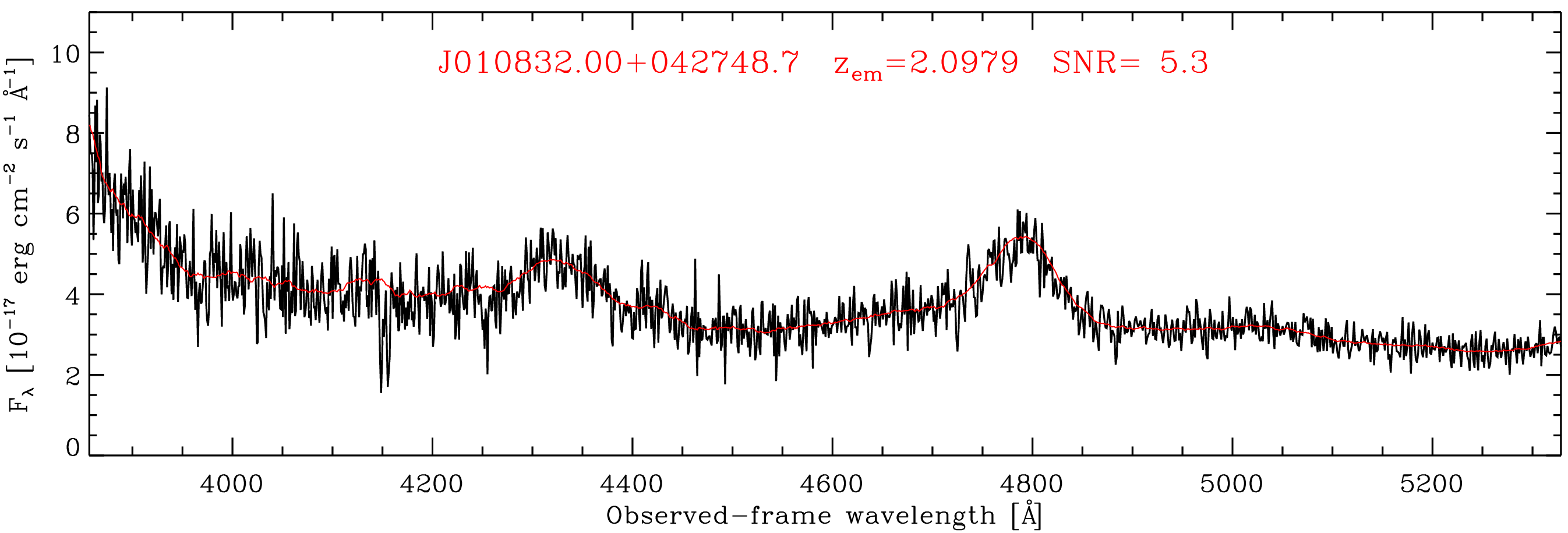}
\hspace{2ex}
\includegraphics[width=7.6cm,height=1.5cm]{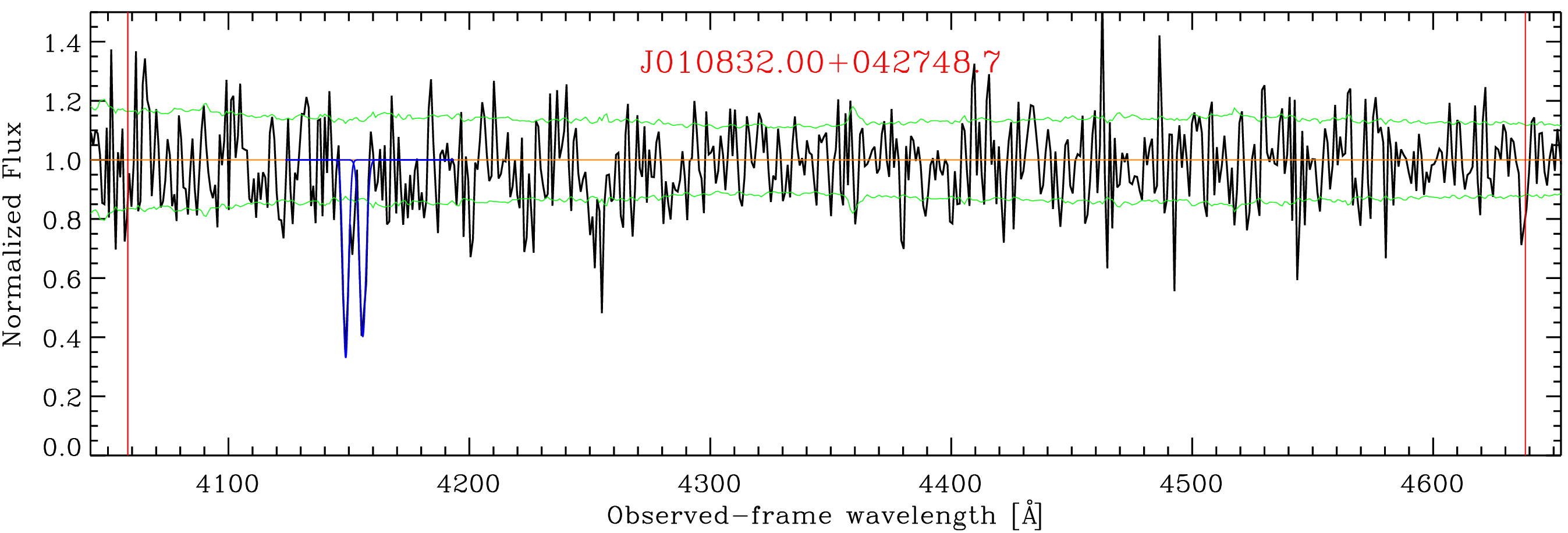}
\includegraphics[width=7.6cm,height=1.5cm]{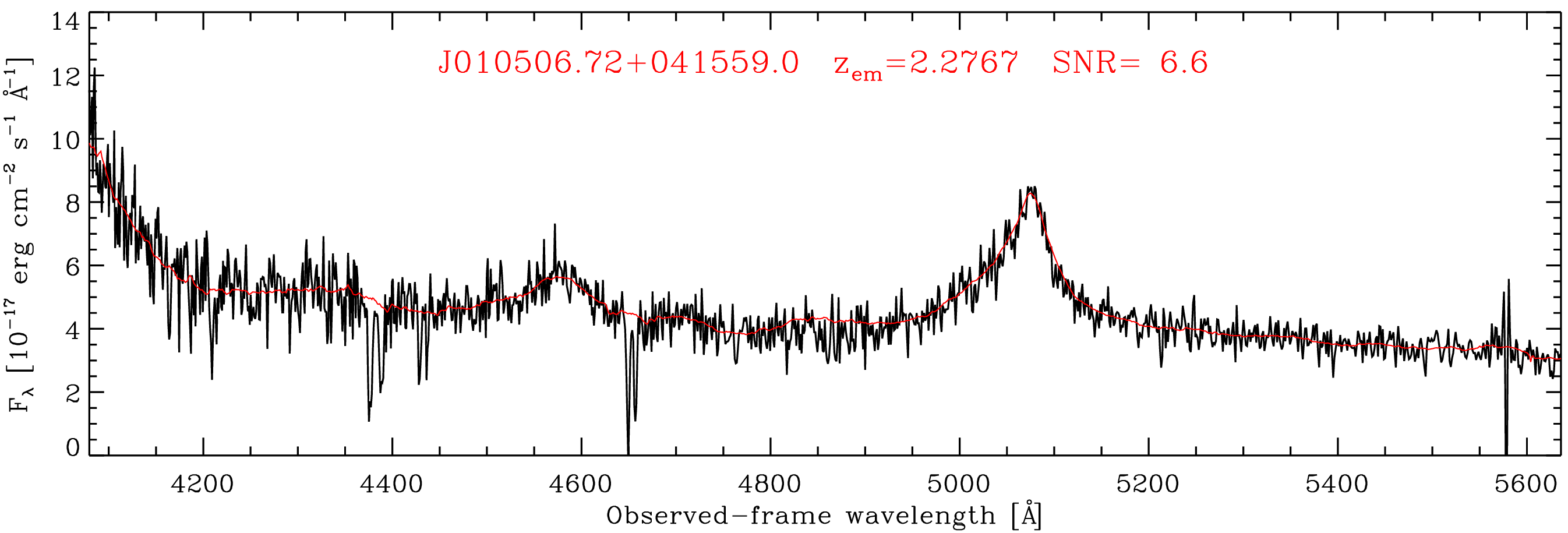}
\hspace{2ex}
\includegraphics[width=7.6cm,height=1.5cm]{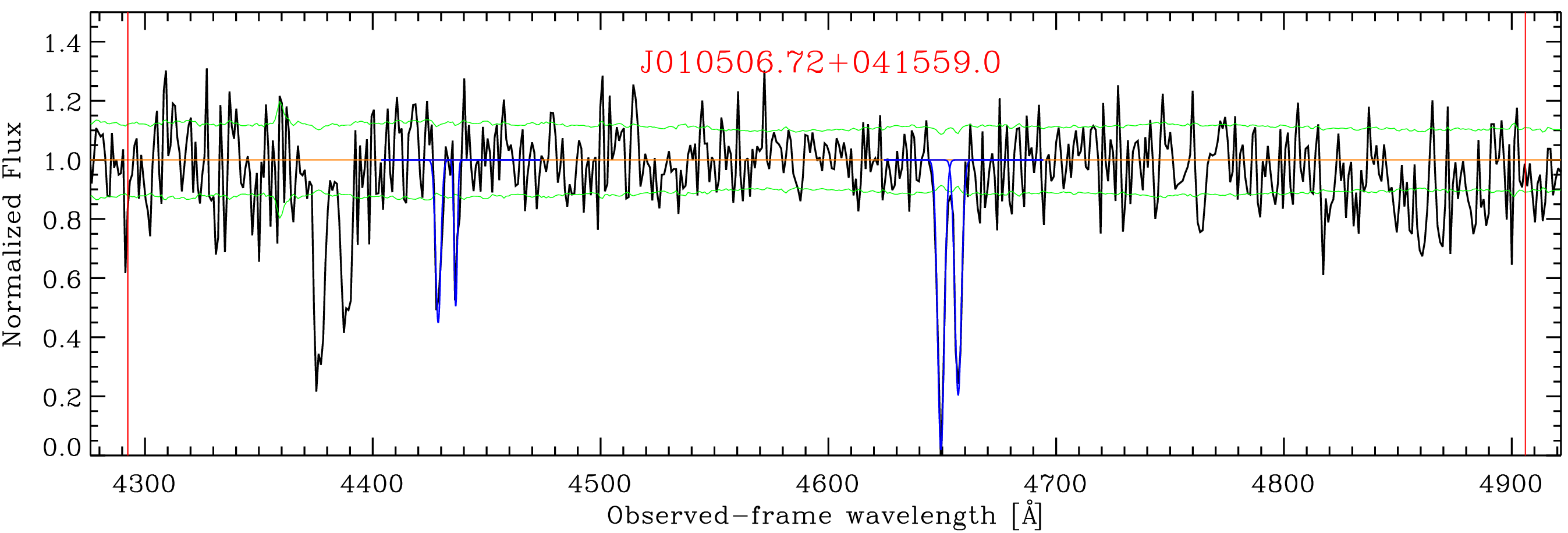}
\includegraphics[width=7.6cm,height=1.5cm]{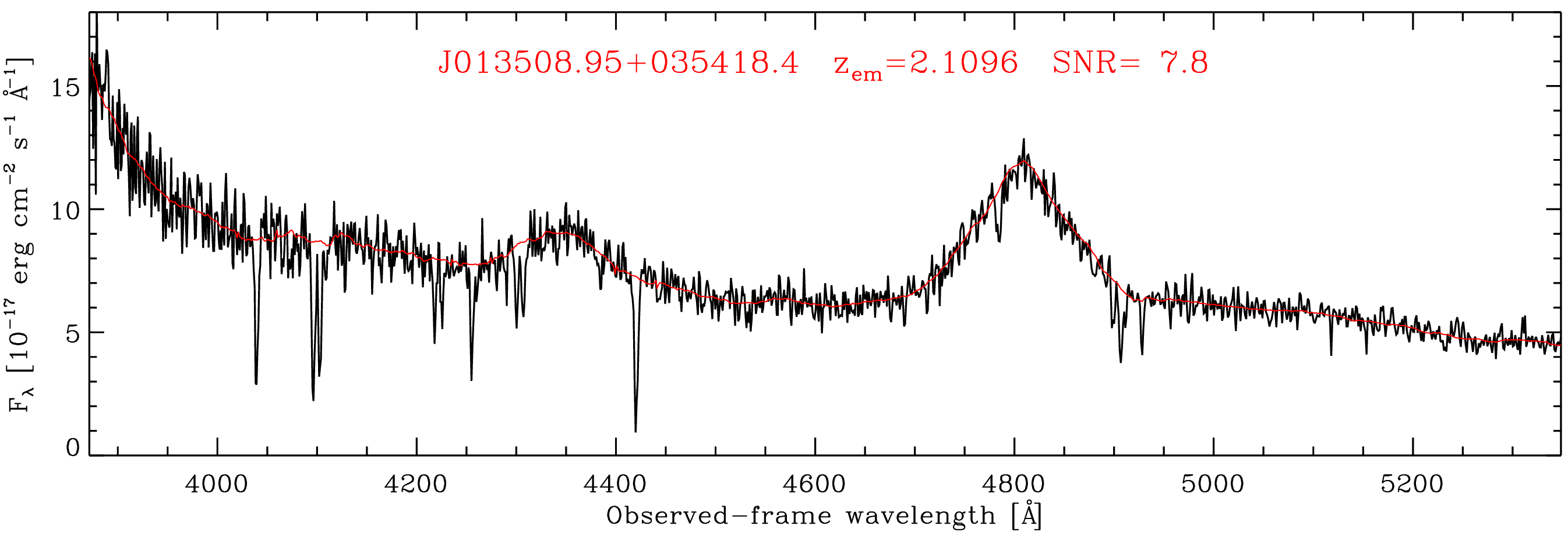}
\hspace{2ex}
\includegraphics[width=7.6cm,height=1.5cm]{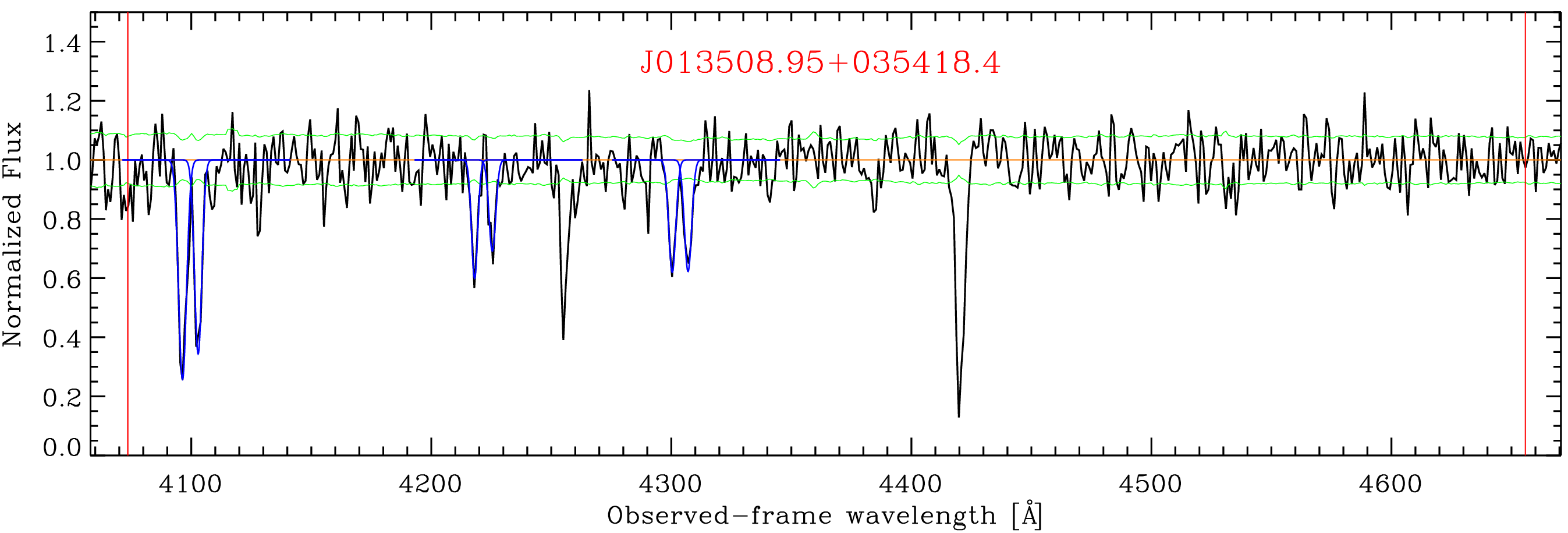}
\includegraphics[width=7.6cm,height=1.5cm]{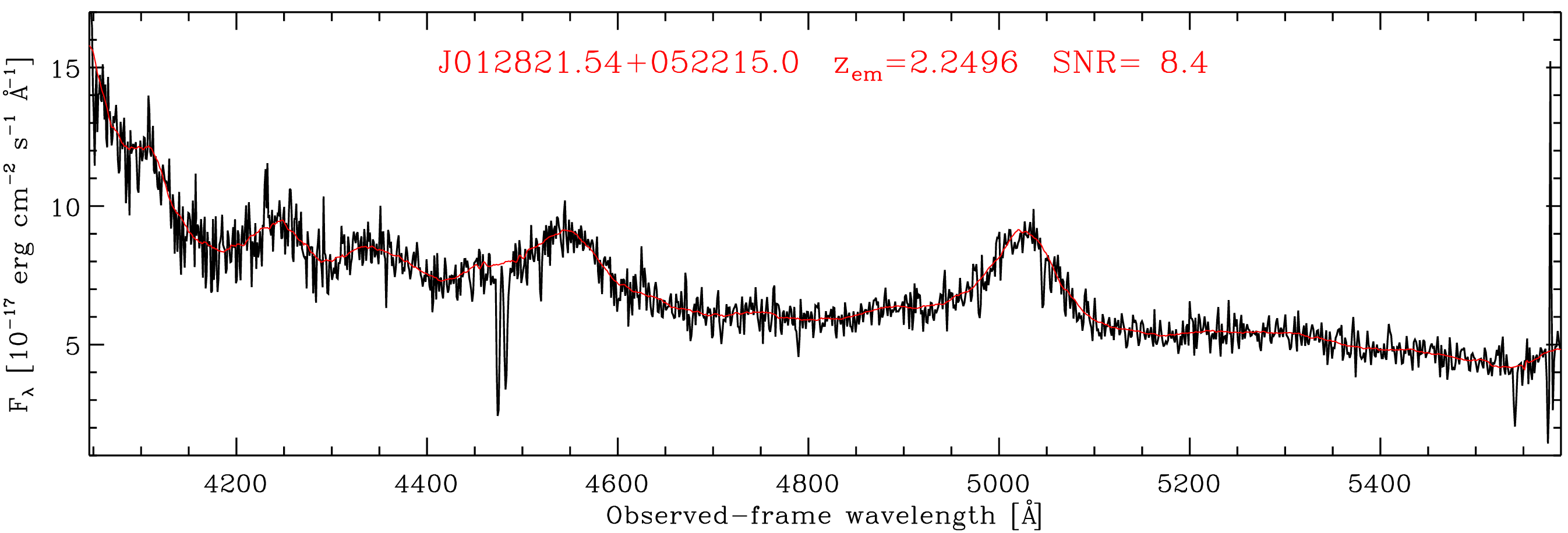}
\hspace{2ex}
\includegraphics[width=7.6cm,height=1.5cm]{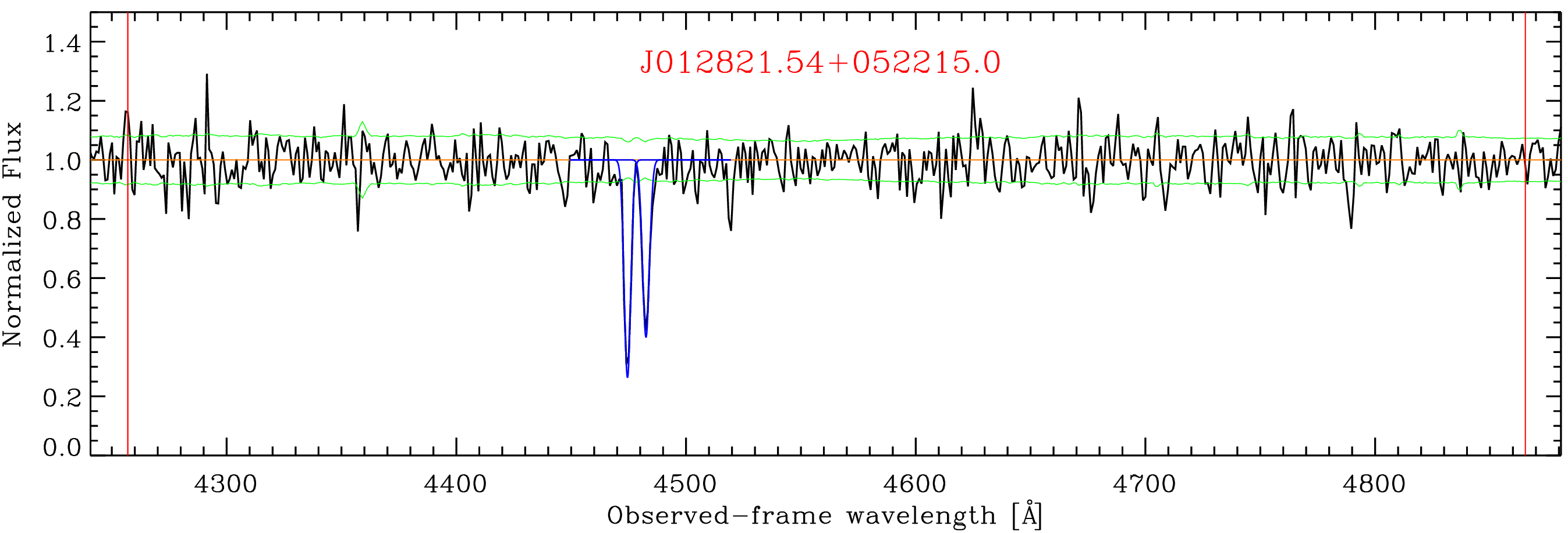}
\includegraphics[width=7.6cm,height=1.5cm]{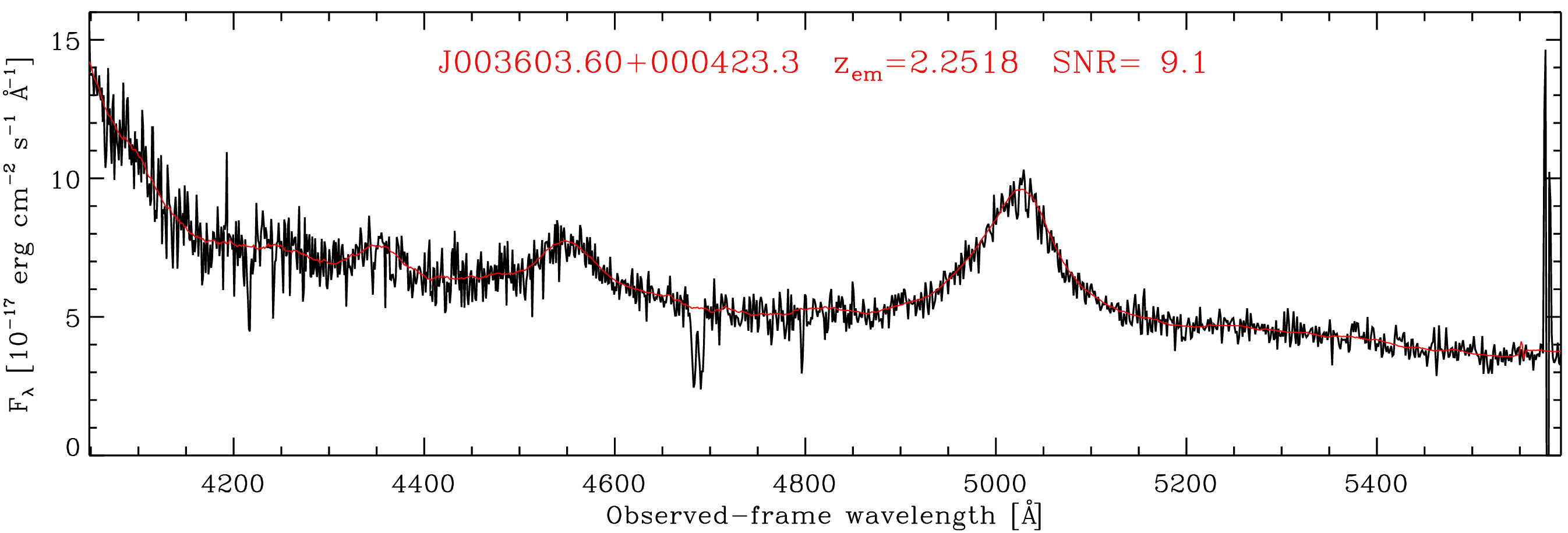}
\hspace{2ex}
\includegraphics[width=7.6cm,height=1.5cm]{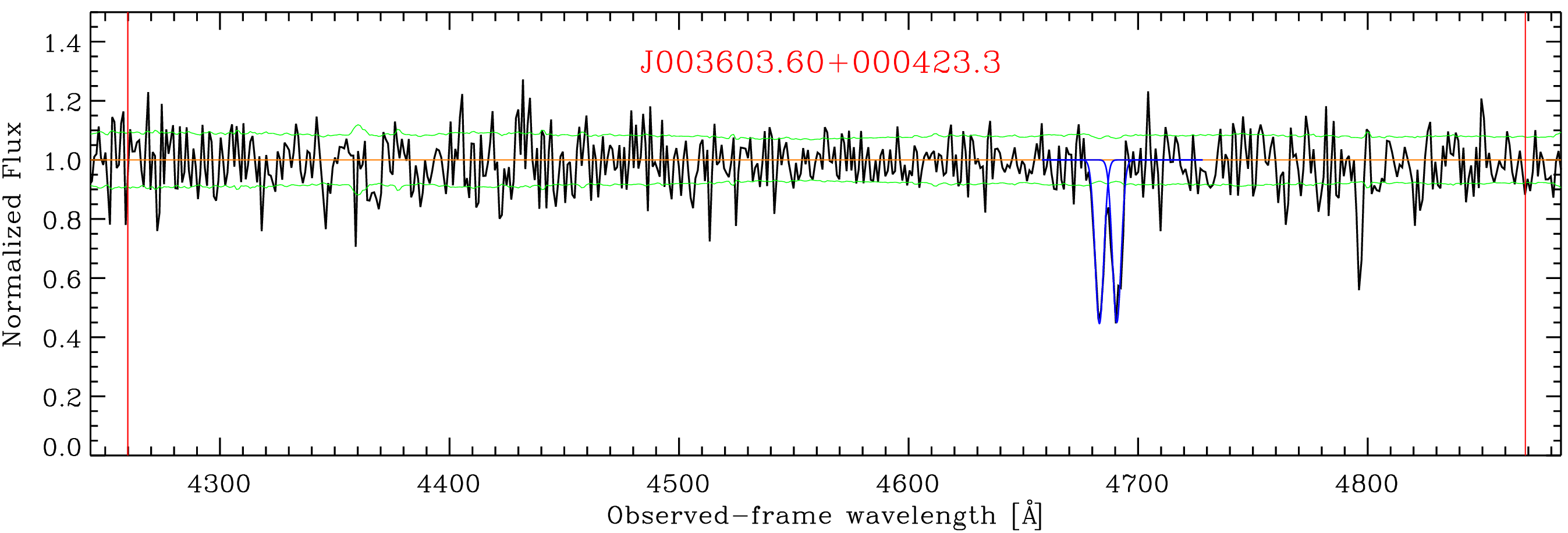}
\includegraphics[width=7.6cm,height=1.5cm]{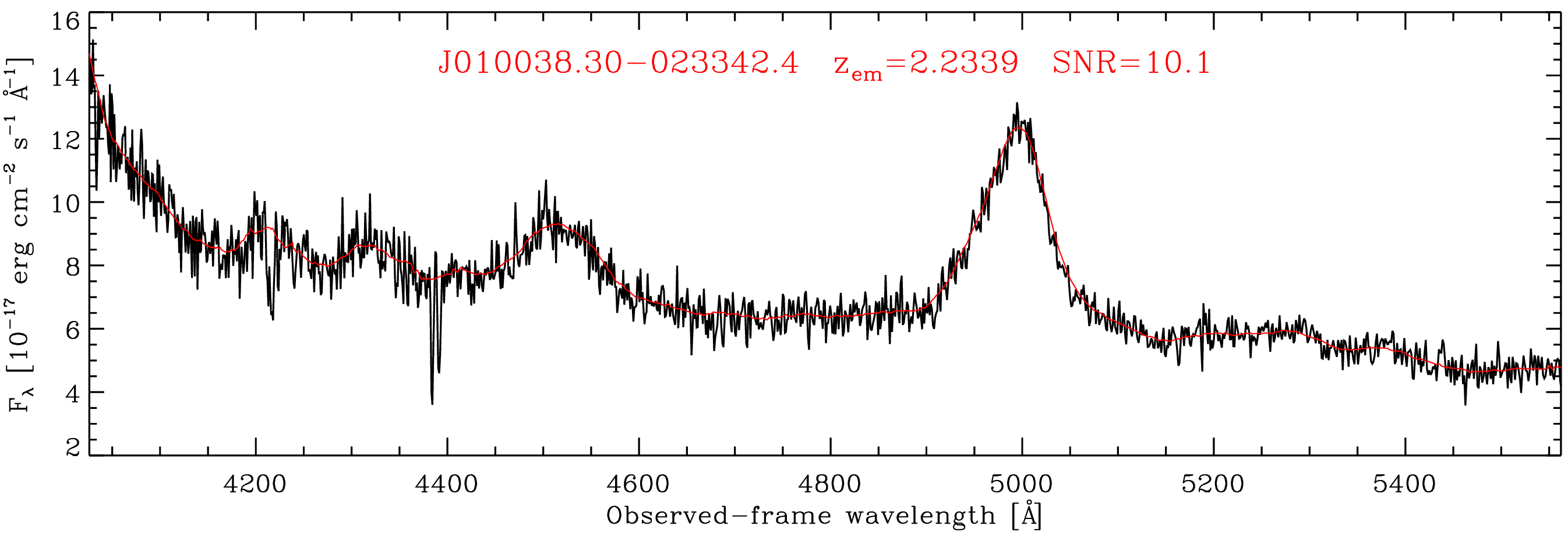}
\hspace{2ex}
\includegraphics[width=7.6cm,height=1.5cm]{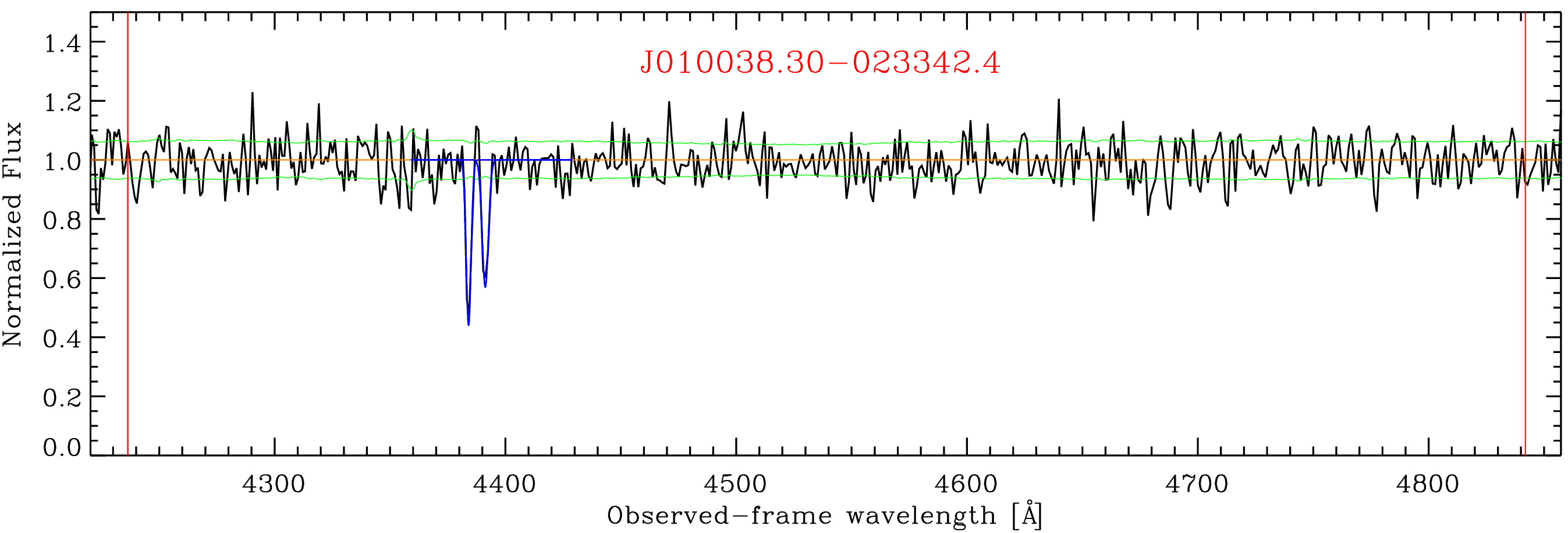}
\includegraphics[width=7.6cm,height=1.5cm]{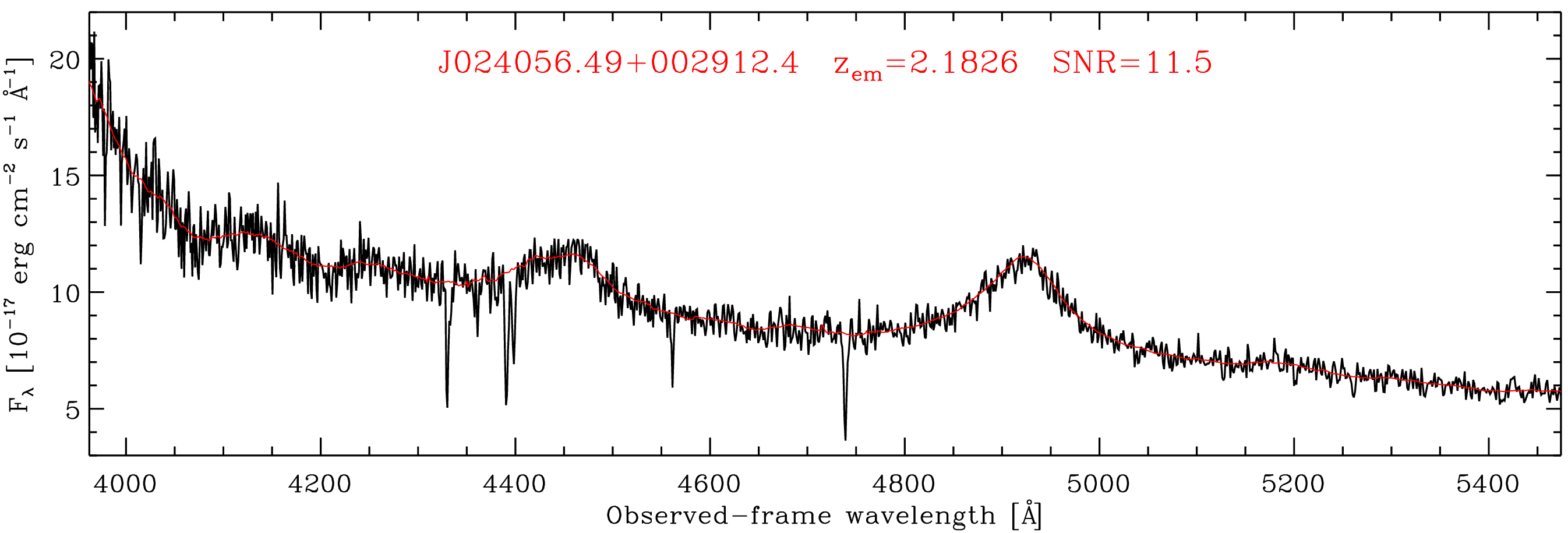}
\hspace{2ex}
\includegraphics[width=7.6cm,height=1.5cm]{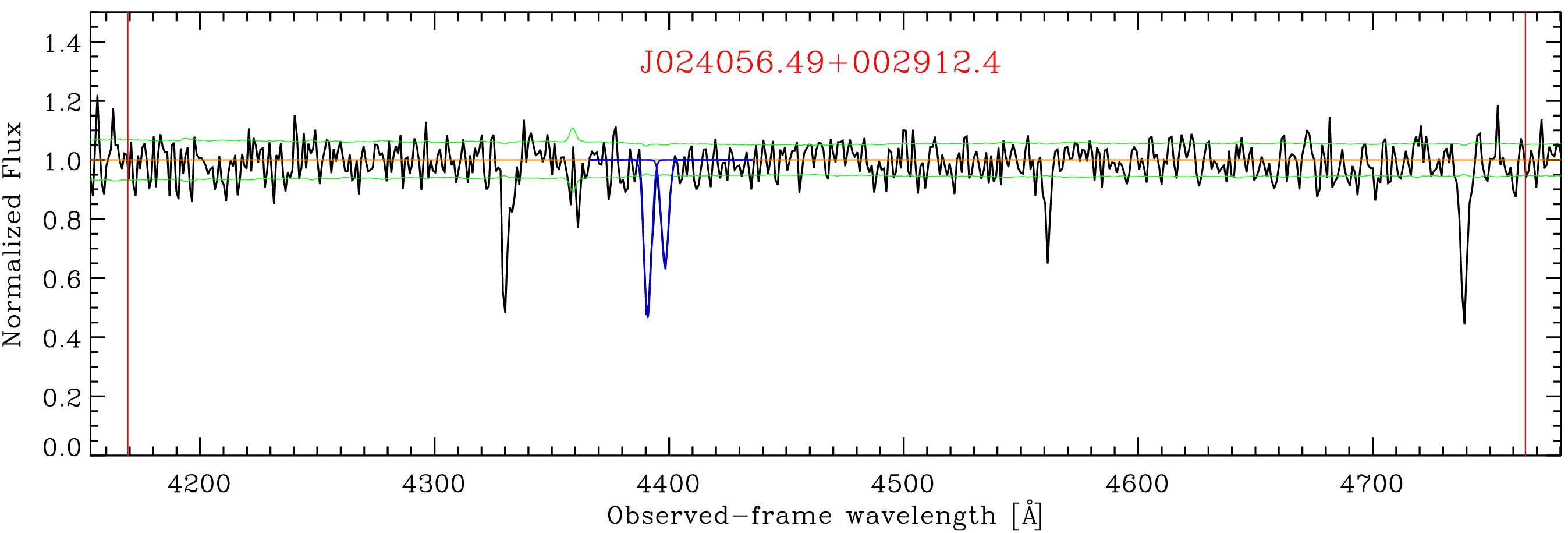}
\includegraphics[width=7.6cm,height=1.5cm]{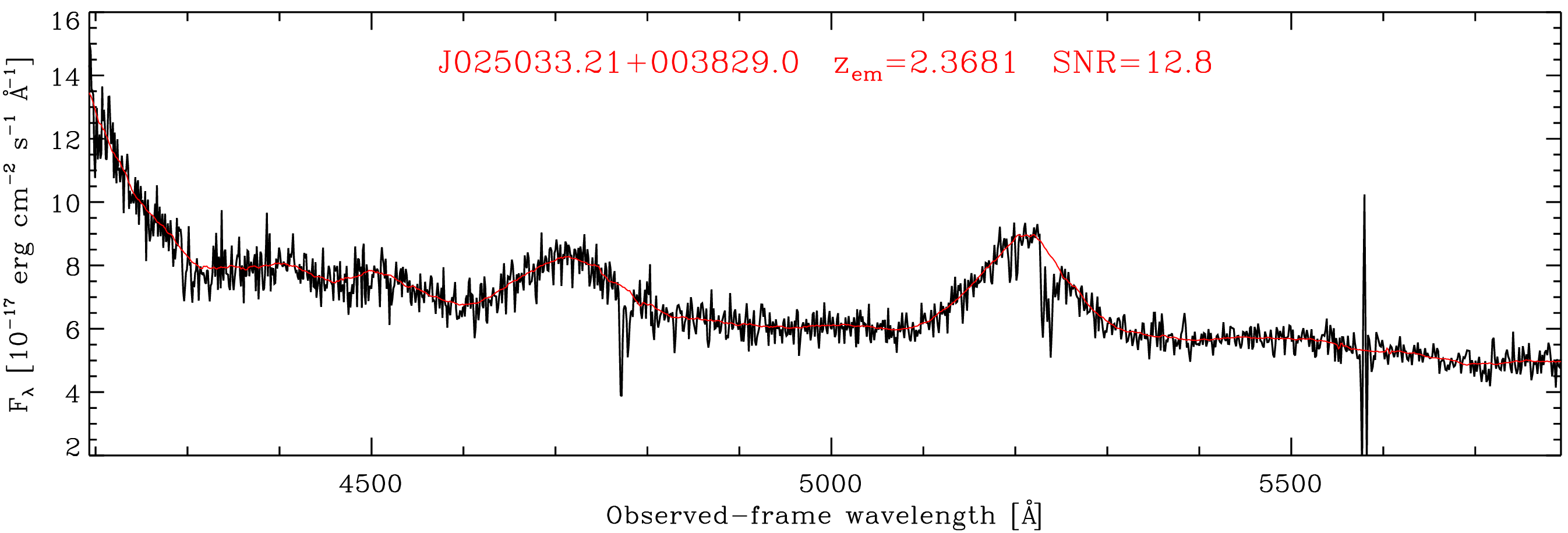}
\hspace{2ex}
\includegraphics[width=7.6cm,height=1.5cm]{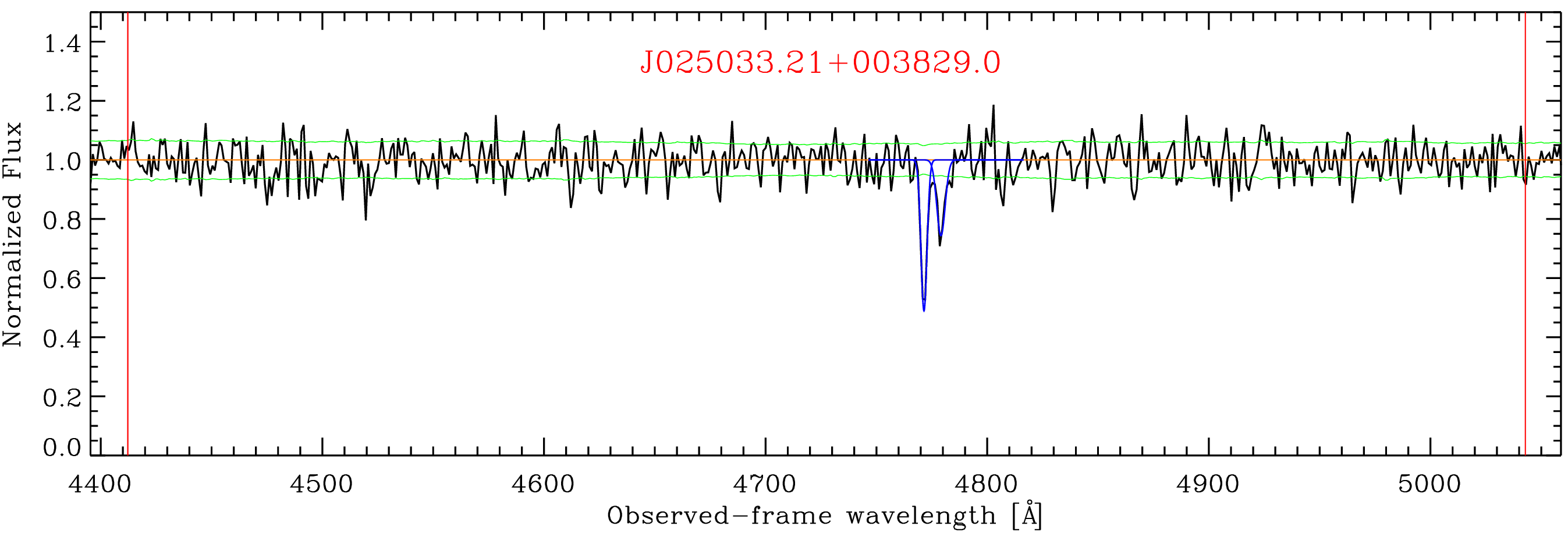}
\includegraphics[width=7.6cm,height=1.5cm]{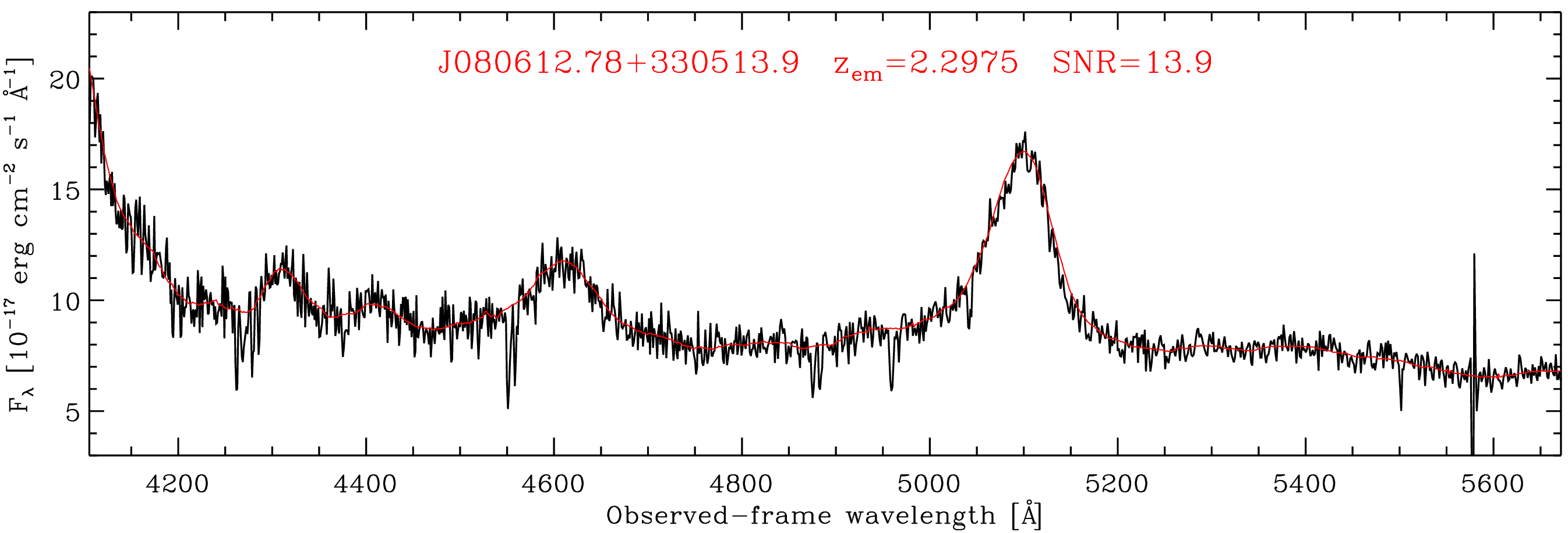}
\hspace{2ex}
\includegraphics[width=7.6cm,height=1.5cm]{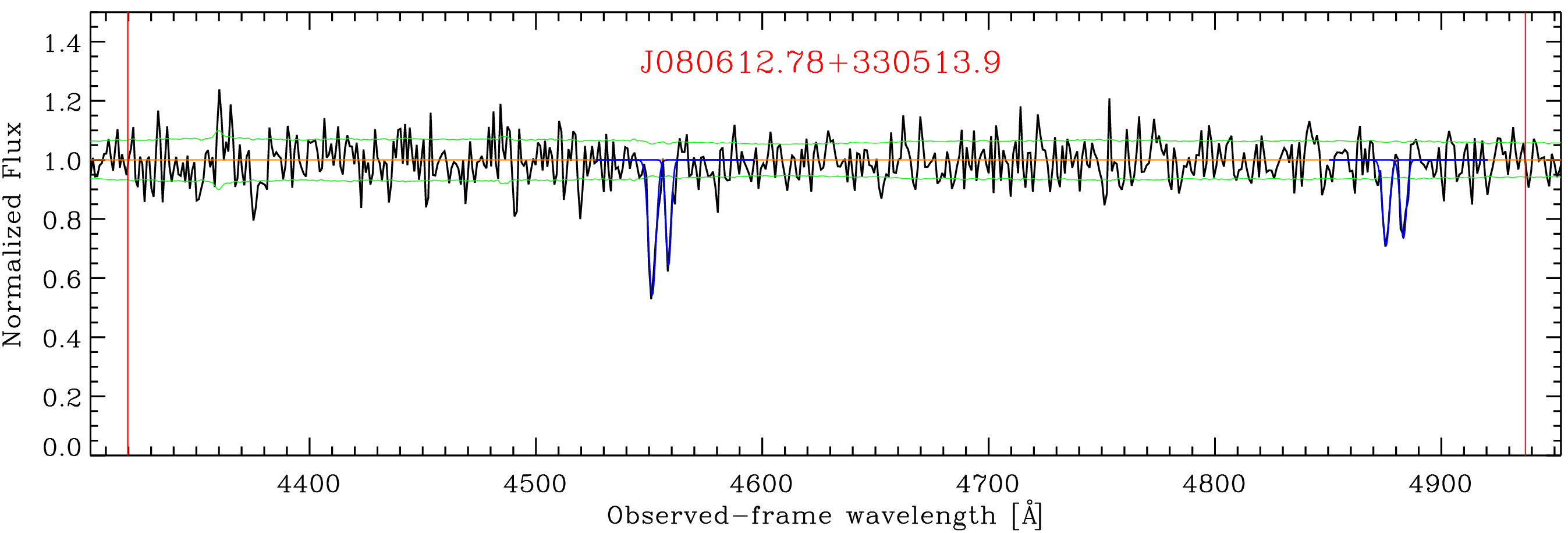}
\includegraphics[width=7.6cm,height=1.5cm]{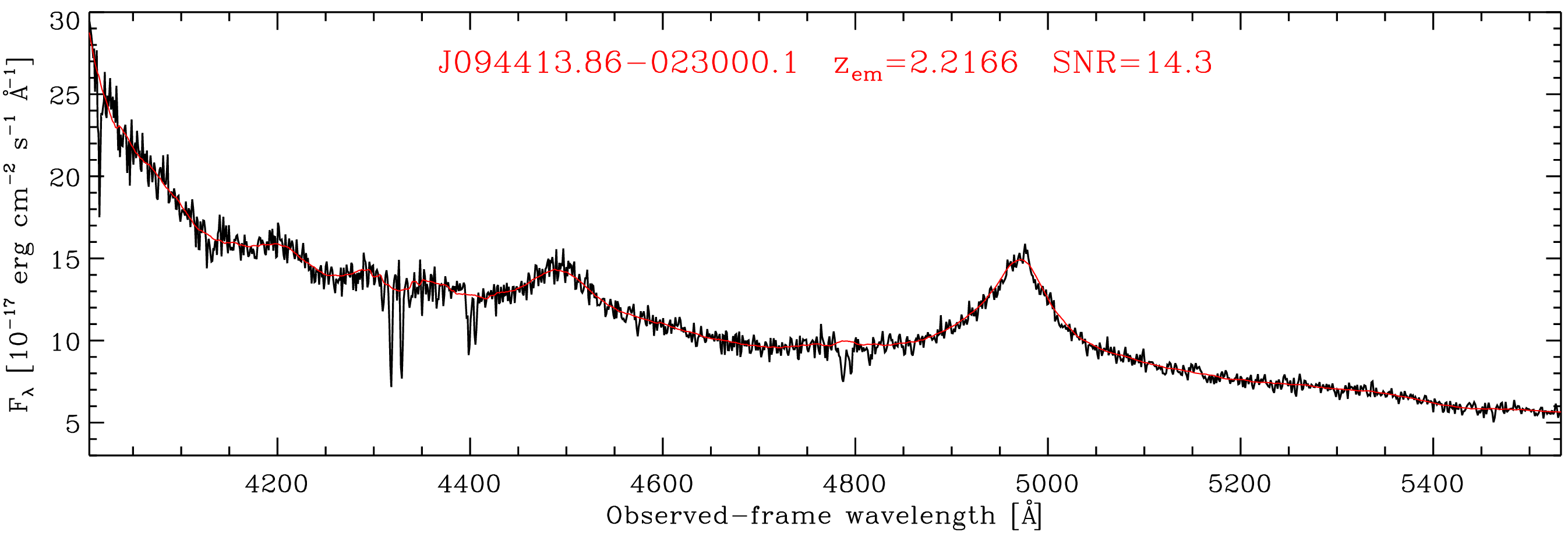}
\hspace{2ex}
\includegraphics[width=7.6cm,height=1.5cm]{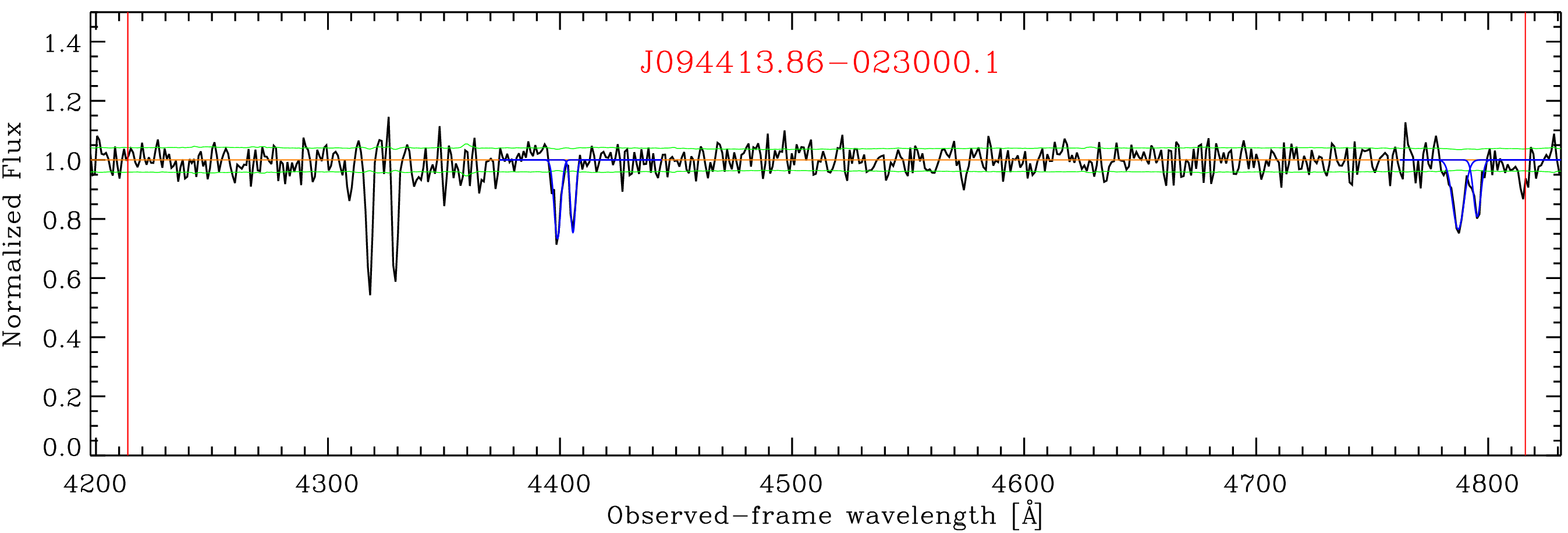}
\caption{The quasar spectra with various values of the median
signal-to-noise ratio (SNR) in the searched spectral region of $\rm
C~IV\lambda\lambda1548,1551$ absorption doublets. The red curves in
the left panels represent the pseudo-continuum fitting curves. The
green lines in the right panels represent the $1\sigma$ flux
uncertainty levels that have been normalized by the
pseudo-continuum. The blue solid lines are the Gaussian fitting
curves of the doublets. The red vertical lines in the right panels
represent the lower and upper limitations respectively, which are
used to cut the spectral region to search for $\rm
C~IV\lambda\lambda1548,1551$ absorption doublets. We do not search
the $\rm C~IV\lambda\lambda1548,1551$ absorption doublet in the
spectra with SNR less than 4 (e.g., the first five spectra).}
\end{figure*}

\begin{table*}[htbp]
\caption{Catalog of $\rm C~IV\lambda\lambda1548,1551$ absorption
systems} \tabcolsep 1.1mm \centering 
 \begin{tabular}{cccccccccccccc}
 \hline\hline\noalign{\smallskip}
SDSS NAME & PLATEID & MJD & FIBERID & $z_{\rm em}$ & $z_{\rm abs}$ &
$\rm W_r\lambda1548$ &$N_{\sigma\lambda1548}$& $\rm W_r\lambda1551$&$N_{\sigma\lambda1551}$&$SNR^{\lambda1548}$&$SNR^{\lambda1551}$&$\beta$\\
\hline\noalign{\smallskip}
000027.01+030715.5  &   4296    &   55499   &   0630    &   2.3533  &   1.9833  &   0.22    &   4.40    &   0.22    &   4.40    &   3.9     &   4.4     &   0.11639     \\
000027.01+030715.5  &   4296    &   55499   &   0630    &   2.3533  &   2.1303  &   0.91    &   22.75   &   0.69    &   17.25   &   20.3    &   18.3    &   0.06871     \\
000050.59+010959.1  &   4216    &   55477   &   0746    &   2.3678  &   1.8971  &   0.46    &   7.67    &   0.47    &   5.88    &   7.1     &   5.3     &   0.14942     \\
000050.59+010959.1  &   4216    &   55477   &   0746    &   2.3678  &   1.9184  &   0.99    &   14.14   &   0.86    &   14.33   &   13.6    &   13.0    &   0.14225     \\
000120.27+030731.9  &   4277    &   55506   &   0098    &   2.1082  &   1.8898  &   0.38    &   7.60    &   0.25    &   6.25    &   6.6     &   5.3     &   0.07273     \\
000133.39+023657.1  &   4277    &   55506   &   0090    &   1.6556  &   1.4773  &   0.71    &   2.84    &   0.67    &   4.47    &   2.7     &   4.2     &   0.06939     \\
000146.95+001428.9  &   4216    &   55477   &   0860    &   2.1567  &   1.9256  &   0.39    &   3.90    &   0.38    &   6.33    &   3.8     &   5.6     &   0.07588     \\
000202.33-002648.4  &   4216    &   55477   &   0154    &   2.1761  &   1.9382  &   0.59    &   3.47    &   0.35    &   3.18    &   3.3     &   2.9     &   0.07770     \\
000207.61+032801.5  &   4296    &   55499   &   0748    &   2.2195  &   1.7502  &   1.25    &   6.58    &   0.86    &   6.14    &   6.2     &   5.9     &   0.15626     \\
000223.32+010101.2  &   4216    &   55477   &   0876    &   2.2931  &   2.1549  &   0.32    &   2.91    &   0.45    &   3.21    &   2.6     &   3.1     &   0.04285     \\
\hline\hline\noalign{\smallskip}
\end{tabular}
\\
 \footnote[]~Note---$N_{\sigma}=\frac{W_r}{\sigma_{W}}$ represents the significant level of the detection.  $\beta=\frac{v}{c}=\frac{(1+z_{em})^2-(1+z_{abs})^2}{(1+z_{em})^2+(1+z_{abs})^2}$.
The table is available in its entirety in the machine-readable form
in the online journal.
\end{table*}

\section{Statistical properties of the absorbers}
In this work, we collect 10,121 quasars to identify $\rm
C~IV\lambda\lambda1548,1551$ absorption doublets, whose emission
redshifts are plotted in Fig. 4. Of the 10,121 quasar spectra, 5,442
are found to have at least one detected $\rm
C~IV\lambda\lambda1548,1551$ absorption doublet. Emission redshifts
of these 5,442 quasars are also plotted in Fig. 4. We identify 8,368
$\rm C~IV\lambda\lambda1548,1551$ absorption doublets from these
quasars. These absorption redshifts are also showed in the Fig. 4.

\begin{figure}
\vspace{3ex}\centering
\includegraphics[width=7 cm,height=6 cm]{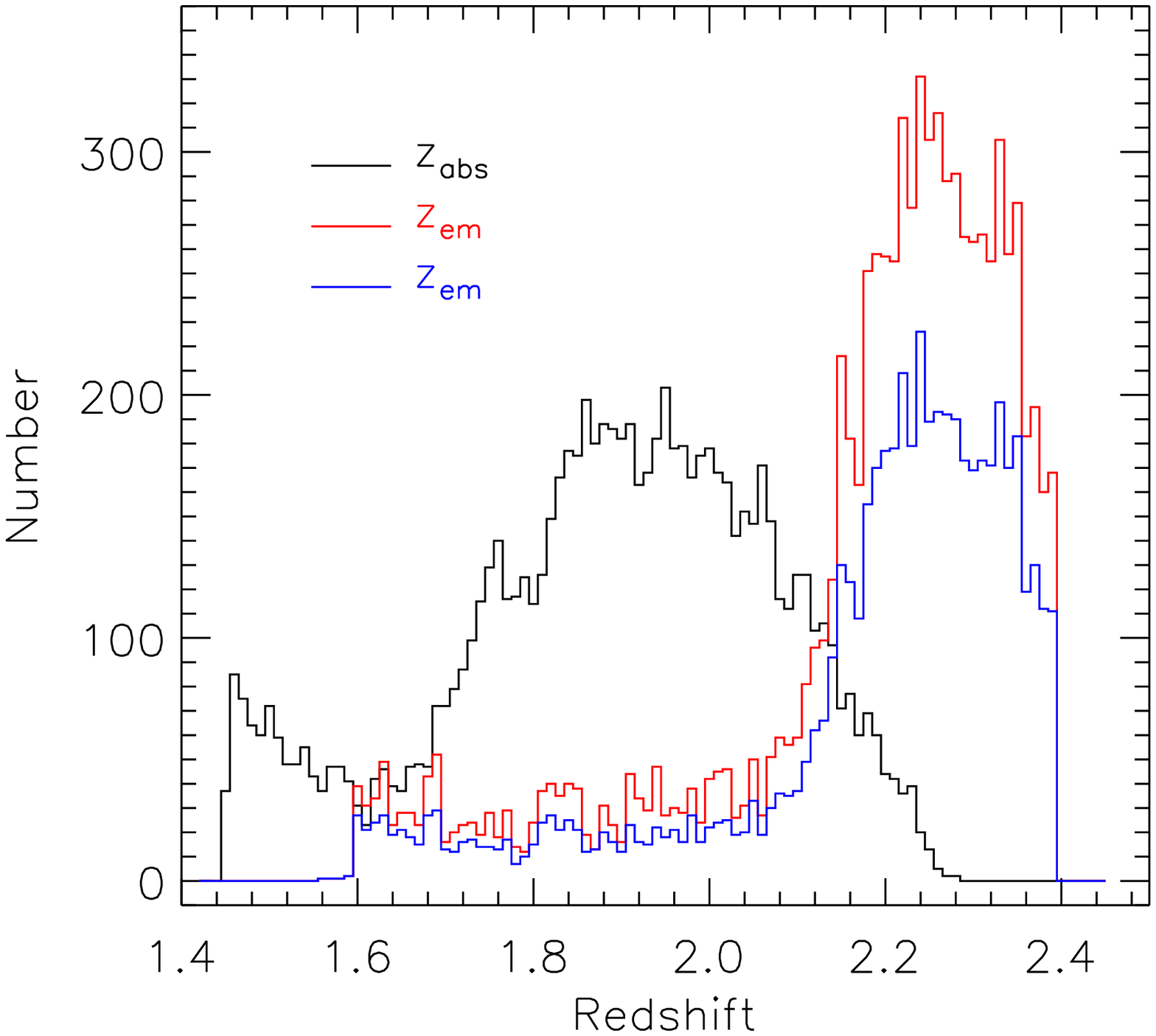}\vspace{3ex}
\caption{Distributions of redshifts. The red line represents the
emission redshift of 10,121 quasars that are used to search for $\rm
C~IV\lambda\lambda1548,1551$ absorption doublets. The blue line
stands for the emission redshift of the 5442 quasars for which at
least one $\rm C~IV\lambda\lambda1548,1551$ absorption doublet is
detected. The black line describes the absorption redshift of all
the detected $\rm C~IV\lambda\lambda1548,1551$ absorption doublets.}
\end{figure}

The total redshift path covered by this catalog can be computed via
\begin{equation}
Z(SNR^{\lambda1548})=\sum_{i=1}^{N_{spec}}\int_{z_i^{min}}^{z_i^{max}}g_i(SNR^{\lambda1548},z)dz,
\end{equation}
where $\rm g_i(SNR^{\lambda1548},z)=1$ if $\rm SNR^{lim}~\le
SNR^{\lambda1548}$, otherwise $\rm g_i(SNR^{\lambda1548},z)=0$; $\rm
z_i^{min}$ and $\rm z_i^{max}$ are the redshifts corresponding to
the minimum and maximum wavelengths of survey for quasar $i$,
respectively (see also Qin et al. 2013). The derived redshift path
is shown in Fig. 5 as a function of the signal-to-noise ratio.

\begin{figure}
\centering
\includegraphics[width=7cm,height=6cm]{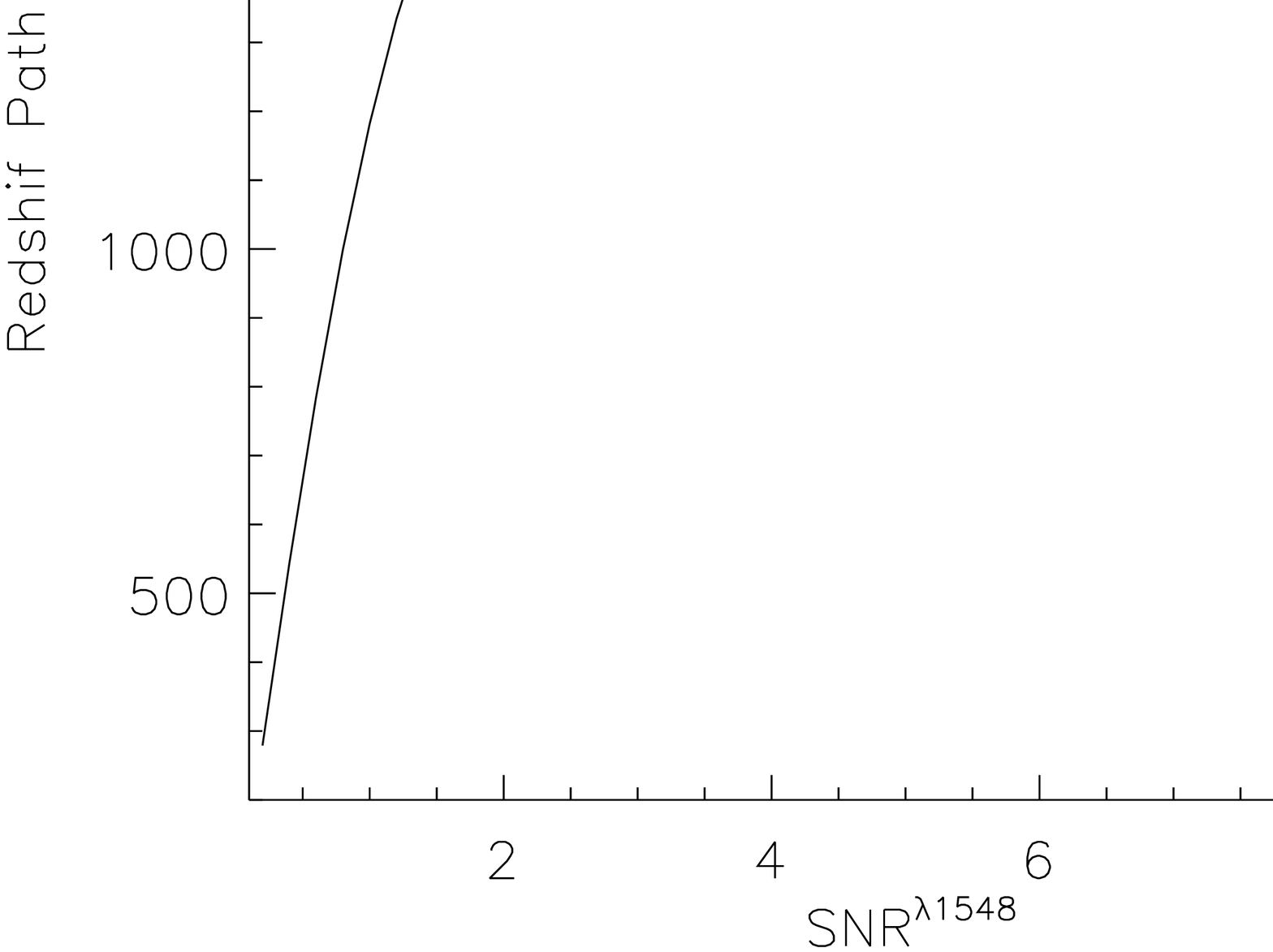}
\caption{Redshift path covered by our catalog ($z_{\rm em} \le
2.4$), shown as a function of $\rm SNR^{\lambda1548}$.}
\end{figure}

Distributions of $W_r$ of the two lines of the $\rm
C~IV\lambda\lambda1548,1551$ absorption doublet of our catalog are
plotted in Fig. 6. These distributions have smooth tails out to
$W_r\approx3.0$ \AA,~ with the largest values of
$W_r\lambda1548=3.19$ \AA~ and $W_r\lambda1551=2.88$ \AA,~
respectively. The median values of the $W_r$ are: 0.62 \AA~ for the
$\lambda1548$ absorption lines, and 0.49 \AA~ for the $\lambda1551$
absorption lines. In this catalog, about 33.7\% (2823/8368)
absorbers have $0.2$ \AA$\le W_r\lambda1548<0.5$ \AA,~ about 45.9\%
(3842/8368) absorbers have $0.5$ \AA$\le W_r\lambda1548<1.0$ \AA,~
about 19.2\% (1603/8368) absorbers have $1.0$ \AA$\le
W_r\lambda1548<2.0$ \AA,~ and about 1.2\% (100/8368) absorbers have
$W_r\lambda1548\ge2.0$ \AA.

\begin{figure}
\vspace{3ex}\centering
\includegraphics[width=7 cm,height=6 cm]{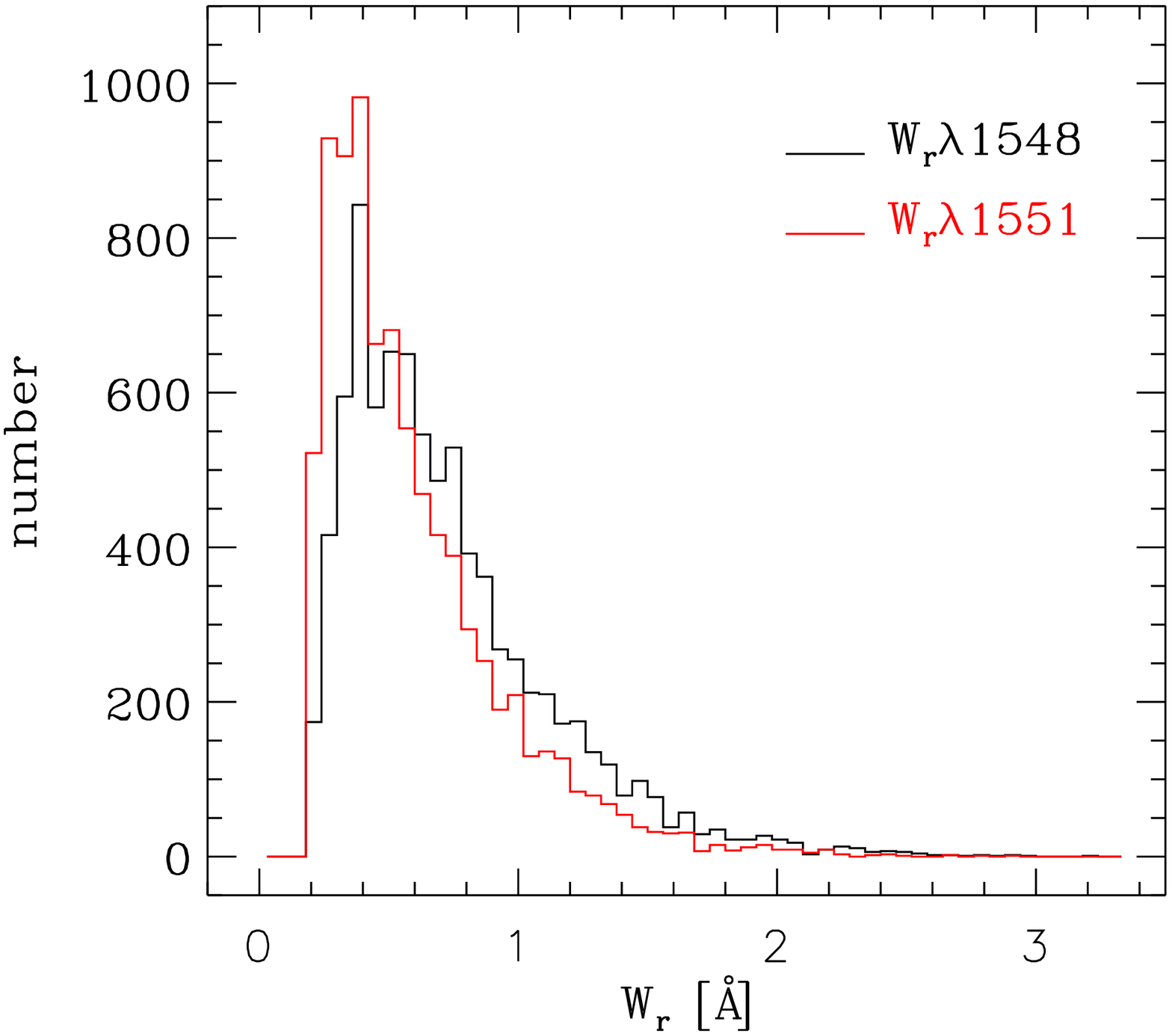}\vspace{3ex}
\caption{Distributions of the rest-frame equivalent width of the
$\rm C~IV$ absorption line. The black line is for the $\rm
\lambda1548$ absorption, and the red line is for the $\rm
\lambda1551$ absorption.}
\end{figure}

In Fig. 7 we plot the distribution of the $W_r$ ratio of the two
lines ($W_r\lambda1548/W_r\lambda1551$). We invoke a Gaussian
function to fit this distribution, which yields a center value of
1.18 and $\rm FWHM=0.80$. The maximum and minimum values of the
$W_r$ ratio are 4.5 and 0.2, respectively. The $W_r$ ratio can
reflects the saturated degree (Str\"omgren 1948). The $W_r$ ratio of
the $\rm C~IV\lambda\lambda1548,1551$ doublet can be changed from
completely saturated absorption, $\rm DR = 1.0$, to completely
unsaturated absorption, $\rm DR = 2.0$ (e.g., Sargent et al. 1988;
Steidel 1990). The boundaries of the completely saturated absorption
($\rm DR = 1.0$) and completely unsaturated absorption ($\rm DR =
2.0$) are marked in Fig. 7. Most of the absorbers of this catalog
satisfy $1.0\le W_r\lambda1548/W_r\lambda1551 \le 2.0$, occupying
nearly 72.9\% (6007/8638) of the total. About 22.0\% (1839/8638)
absorbers have $W_r\lambda1548/W_r\lambda1551<1.0$, and about 6.2\%
(522/8638) absorbers have $W_r\lambda1548/W_r\lambda1551>2.0$. We
guess that the $\rm C~IV\lambda\lambda1548,1551$ absorption systems
that lie outside the theoretical limits of the $W_r$ ratio
($W_r\lambda1548/W_r\lambda1551<1.0$ or
$W_r\lambda1548/W_r\lambda1551>2.0$) might mainly originate from the
line blending.

\begin{figure}
\vspace{3ex}\centering
\includegraphics[width=7.5 cm,height=6.5 cm]{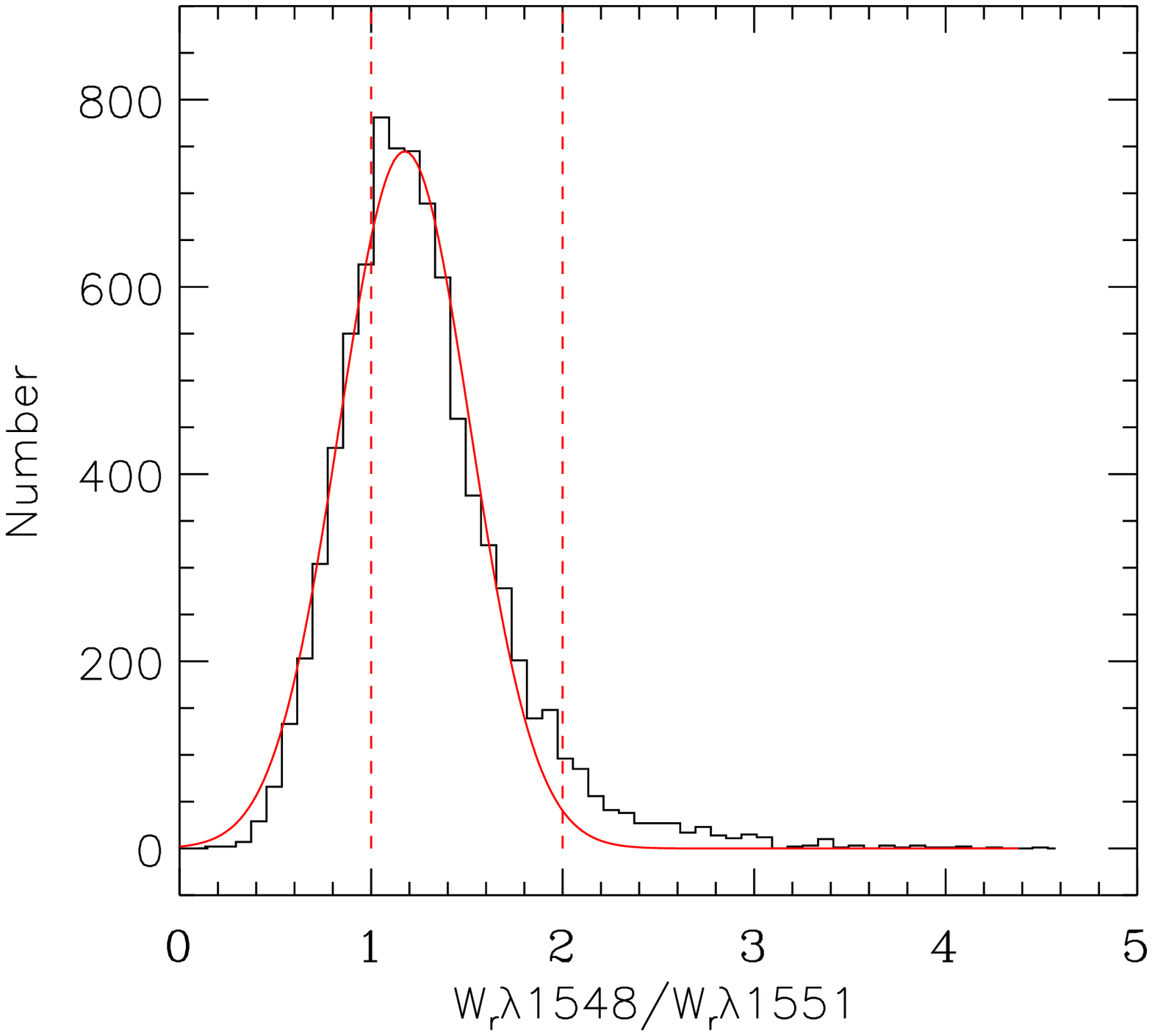}\vspace{3ex}
\caption{Distribution of the ratio of the rest-frame equivalent
widthes of the $\rm C~IV$ doublet. The red curve is the fitting
Gaussian with $\rm center = 1.18~and~FWHM =0.80$. The red dash lines
are the theoretical limits for completely saturated ($\rm
W_r\lambda1548/W_r\lambda1551 = 1.0$) and unsaturated ($\rm
W_r\lambda1548/W_r\lambda1551 = 2.0$) absorptions, respectively.}
\end{figure}

\section{Discussion}
In order to estimate the false positives/negatives of the $\rm C~IV$
absorption system, we wish to look at the frequency of the detected
$\rm C~IV$ absorption systems ($f_{NALs}$) as a function of
signal-to-noise ratio, which can be computed via
\begin{equation}
f_{NALs}=\lim_{\bigtriangleup SNR\rightarrow0}\frac{\bigtriangleup
N_{abs}}{\bigtriangleup N_{sdp}}
\end{equation}
where $\bigtriangleup N_{abs}$ and $\bigtriangleup N_{sdp}$ are the
count of the detected $\rm C~IV$ absorption systems and the count of
the spectral data points in signal-to-noise ratio bin
$\bigtriangleup SNR$, respectively.
The resulting $f_{NALs}$, as a function of the signal-to-noise
ratio, is displayed in Fig. 8. It exhibits a platform in the range
of $SNR^{\lambda1548}\gtrsim4$, suggesting that the detection of
$C~IV$ absorption systems would likely be complete when the
signal-to-noise ratio is larger than 4.

\begin{figure}
\centering
\includegraphics[width=7cm,height=6cm]{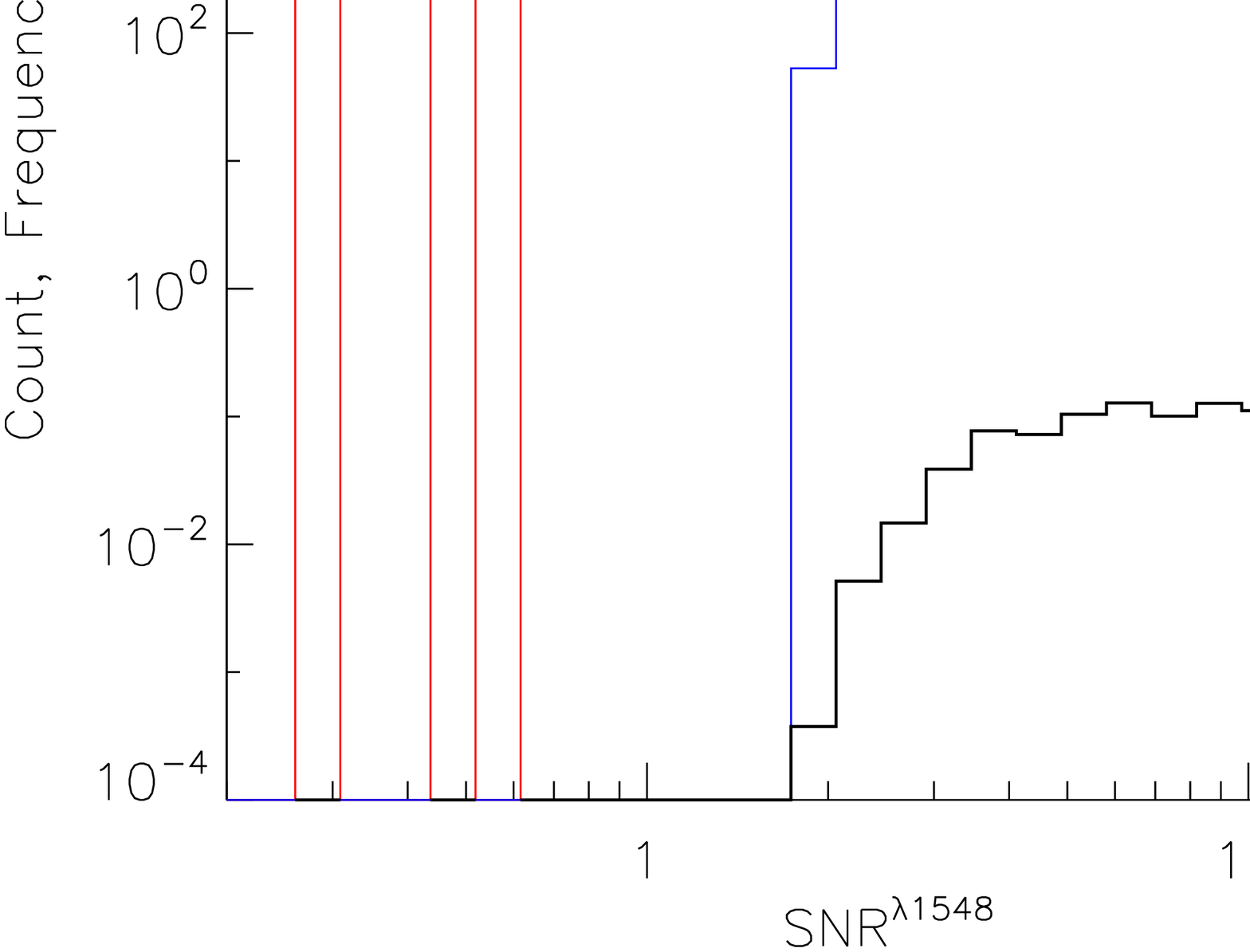}
\caption{Plot of the detection frequency as a function of the
signal-to-noise ratio. The upper (red) line represents the count of
the spectral data point, the middle (blue) line represents the count
of the detected $\rm C~IV$ absorption system, and the bottom (black)
line stands for the frequency of NALs detected in this work,
calculated by equation (5).}
\end{figure}

The incompleteness of the detection of $C~IV$ absorption systems is
obvious within the range of $SNR^{\lambda1548}\lesssim4$. As
suggested by Fig. 8, we find that, within the range of
$SNR^{\lambda1548}\lesssim4$, when the signal-to-noise ratio tends
to be smaller, more $C~IV$ absorption systems would tend to be
missed by our analysis. To roughly estimate the significance of the
incompleteness, we compute the missing rate ($f_{MR}$) of the
detection of $C~IV$ absorption systems in several bins of the
signal-to-noise ratio via
\begin{equation}
f_{MR}=\frac{\overline{f_{NALs}}-f_{NALs}}{\overline{f_{NALs}}},
\end{equation}
where $\overline{f_{NALs}}$ is the average frequency of NALs in the
range of $SNR^{\lambda1548}>4$, and $f_{NALs}$ is the frequency of
NALs in the corresponding signal-to-noise ratio bin. The results are
presented in Table 2.

\begin{table}
\caption{The missing rate of absorption systems with $SNR^{\lambda1548}\le 4$} \tabcolsep 2mm \centering 
 \begin{tabular}{ccccccc}
 \hline\hline\noalign{\smallskip}
SNR bin & [2.0,2.5] & [2.5,3.0] & [3.0,3.5] & [3.5,4.0]\\
\hline\noalign{\smallskip}
$f_{MR}$&0.91&0.67&0.62&0.20\\
\hline\hline\noalign{\smallskip}
\end{tabular}
\end{table}

To refine the quasar sample to search the $\rm
C~IV\lambda\lambda1548,1551$ absorption system, we perform our
analysis under the condition that the spectra examined must have a
median signal-to-noise ratio greater than or equal to $4$. It is
possible that some $\rm C~IV\lambda\lambda1548,1551$ absorption
doublets, which satisfy our criteria of selecting absorption lines,
may be imprinted in the spectra with the median signal-to-noise
ratio being less than $4$, and they will be missed.

To have a look at these possibly missed doublets, we randomly select
$100$ quasars from those located in the left lower region of Fig. 2
(below the horizontal red line and on the left hand side of the
vertical red line), to detect $\rm C~IV\lambda\lambda1548,1551$
absorption doublets with the same criteria described in section 2.
These quasars are listed in Table 3. $15$ $\rm
C~IV\lambda\lambda1548,1551$ absorption doublets are detected from
these quasar spectra, which are presented in Table 4. For this
randomly selected quasar sample, the redshift path computed using
Equation (4) and the frequency of NALs calculated by Equation (5)
are displayed in Figs. 9 and 10, respectively.

\begin{table}
\caption{Sources of the randomly selected quasar sample} \tabcolsep 1.1mm \centering 
 \begin{tabular}{ccccccccc}
 \hline\hline\noalign{\smallskip}
SDSS NAME & PLATEID & MJD & FIBERID & $z_{\rm em}$ & SNR\\
\hline\noalign{\smallskip}
000525.86+030813.5  &   4296    &   55499   &   0908    &   2.1802  &   3.4 \\
00063.085+031327.1  &   4296    &   55499   &   0962    &   2.3788  &   3.9 \\
002059.05+030633.3  &   4300    &   55528   &   0716    &   2.1935  &   3.4 \\
004616.50+011343.0  &   3589    &   55186   &   0864    &   2.1632  &   1.5 \\
005623.89+021253.2  &   4308    &   55565   &   0740    &   2.2631  &   2.1 \\
010618.39+101247.8  &   4551    &   55569   &   0598    &   2.2872  &   1.3 \\
011927.05+000008.0  &   4227    &   55481   &   0036    &   2.3571  &   1.7 \\
013752.51+102410.6  &   4548    &   55565   &   0802    &   2.1453  &   2.7 \\
\hline\hline\noalign{\smallskip}
\end{tabular}
\\
 \footnote[]~Note---SNR is the median signal-to-noise ratio of the quasar in the surveyed spectral
region. The table is available in its entirety in the
machine-readable form in the online journal.
\end{table}

\begin{table*}
\caption{The $\rm C~IV\lambda\lambda1548,1551$ absorption systems of the randomly selected quasar sample} \tabcolsep 1mm \centering 
 \begin{tabular}{cccccccccccccc}
 \hline\hline\noalign{\smallskip}
SDSS NAME & PLATEID & MJD & FIBERIN & $z_{\rm em}$ & $z_{\rm abs}$ &
$\rm W_r\lambda1548$ &$N_{\sigma\lambda1548}$& $\rm W_r\lambda1551$&$N_{\sigma\lambda1551}$& $SNR^{\lambda1548}$&$SNR^{\lambda1551}$&$\beta$\\
\hline\noalign{\smallskip}
075343.86+182204.9  &   4490    &   55629   &   0734    &   2.1708  &   1.9708  &   0.38    &   2.53    &   0.56    &   2.55    &   2.3     &   2.4     &   0.06506     \\
114931.76+360338.8  &   4653    &   55622   &   0042    &   2.2658  &   1.7910  &   0.79    &   2.39    &   1.21    &   3.67    &   2.2     &   3.4     &   0.15582     \\
014848.55+145729.2  &   4658    &   55592   &   0948    &   2.1370  &   1.8690  &   0.31    &   2.38    &   0.44    &   2.44    &   2.2     &   2.3     &   0.08907     \\
152155.41+310942.3  &   4719    &   55736   &   0322    &   2.1108  &   1.8249  &   0.88    &   5.18    &   0.39    &   2.79    &   4.4     &   2.7     &   0.09611     \\
080345.70+422136.2  &   3683    &   55178   &   0178    &   2.0877  &   1.6675  &   0.61    &   3.81    &   0.71    &   3.23    &   3.5     &   3.0     &   0.14525     \\
155717.07+163309.6  &   3922    &   55333   &   0594    &   2.3355  &   2.1045  &   0.81    &   3.12    &   0.71    &   2.84    &   2.9     &   2.7     &   0.07165     \\
081937.46+302718.3  &   4447    &   55542   &   0070    &   2.2037  &   2.0069  &   0.58    &   2.76    &   0.65    &   2.83    &   2.7     &   2.7     &   0.06331     \\
074256.10+481730.0  &   3675    &   55183   &   0520    &   2.2775  &   1.9637  &   1.14    &   3.93    &   0.99    &   3.41    &   3.6     &   3.2     &   0.10030     \\
150553.69+304300.5  &   3876    &   55245   &   0264    &   2.2329  &   1.7556  &   1.02    &   2.83    &   0.74    &   3.08    &   2.7     &   3.0     &   0.15840     \\
134259.55+340404.9  &   3856    &   55269   &   0612    &   2.2972  &   2.1021  &   0.77    &   2.75    &   0.43    &   2.39    &   2.6     &   2.3     &   0.06092     \\
024842.21-000302.1  &   4241    &   55450   &   0265    &   2.0587  &   1.8580  &   0.59    &   3.11    &   0.46    &   2.88    &   3.0     &   2.7     &   0.06776     \\
150914.76+230044.0  &   3962    &   55660   &   0610    &   2.1652  &   1.9392  &   1.04    &   2.89    &   0.79    &   2.39    &   2.7     &   2.3     &   0.07394     \\
150539.79+062612.3  &   4856    &   55712   &   0230    &   2.3698  &   2.0359  &   0.88    &   3.03    &   1.19    &   3.31    &   2.9     &   3.1     &   0.10397     \\
104647.31+382734.8  &   4634    &   55626   &   0932    &   2.2235  &   1.9783  &   1.20    &   2.86    &   0.76    &   2.38    &   2.7     &   2.2     &   0.07895     \\
094705.52+434013.8  &   4569    &   55631   &   0764    &   2.1984  &   1.9500  &   1.14    &   4.75    &   1.60    &   3.48    &   4.1     &   3.3     &   0.08067     \\
\hline\hline\noalign{\smallskip}
\end{tabular}
\\
 \footnote[]~Note---See Table 1 for the meanings of each column.
 \vspace{6ex}
\end{table*}

\begin{figure}
\centering
\includegraphics[width=7cm,height=6cm]{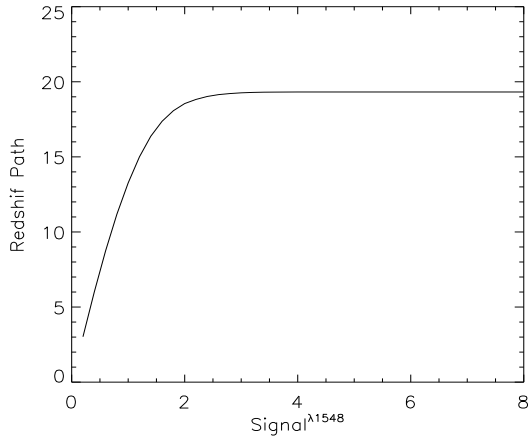}
\caption{Redshift path covered by the randomly selected quasar
sample, shown as a function of $\rm SNR^{\lambda1548}$.}
\end{figure}

\begin{figure}
\centering
\includegraphics[width=7cm,height=6cm]{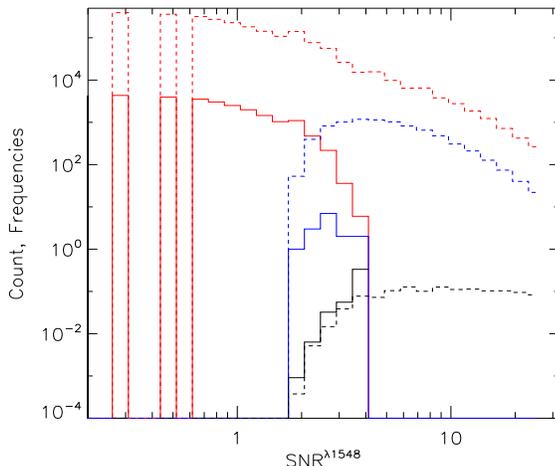}
\caption{Plot of the detection frequency as a function of the
signal-to-noise ratio for the randomly selected quasar sample. See
Fig. 8 for the meanings of each solid line. The dash lines are the
solid lines shown in Fig. 8 with the same colors.}
\end{figure}

The spectral signal-to-noise ratio is important to detect narrow
absorption lines. It is very difficult to distinguish the true NALs
from the noise in the spectra with lower signal-to-noise ratio,
since the fluctuations of the noise frequently confuse or cover the
real narrow absorption lines. As stated above, only 15 $\rm C~IV$
absorption systems are detected in the spectra of the 100 randomly
selected quasars. In other words, only 0.15 $\rm C~IV$ absorption
system can be detected in per quasar spectrum with the median
signal-to-noise ratio being as low as less than 4. However, we
detect 8,368 $\rm C~IV$ absorption systems in the spectra of the
10,121 quasars with their median signal-to-noise ratios being
greater than 4. The value of 8368/10121 is several times larger than
that of 15/100, which manifests that many real absorption lines
cannot be identified in the spectra with lower signal-to-noise
ratios.

\section{Summary}
As the first effort in our series work on identifying absorption
lines in quasar spectra of BOSS, we search quasars with $z_{\rm
em}\le2.4$ and identify potential intervening $\rm
C~IV\lambda\lambda1548,1551$ absorption doublets with
$W_r\lambda1548\ge0.2$ \AA~. Our sample contains 10,121 quasars,
from which we identify 8,368 $\rm C~IV\lambda\lambda1548,1551$
absorption systems which covers the absorption redshift range of
$z_{\rm abs}=1.4544$
--- $2.2805$. Of 10,121 quasars, 5,442 are
detected to have at least one $\rm C~IV\lambda\lambda1548,1551$
absorption doublet. We find that about 33.7\% absorbers have $0.2$
\AA$\le W_r\lambda1548<0.5$ \AA,~ about 45.9\% absorbers have $0.5$
\AA$\le W_r\lambda1548<1.0$ \AA,~ about 19.2\% absorbers have $1.0$
\AA$\le W_r\lambda1548<2.0$ \AA,~ and about 1.2\% absorbers have
$W_r\lambda1548\ge2.0$ \AA. Most of the $\rm
C~IV\lambda\lambda1548,1551$ absorption doublets (72.9\%) lie within
the theoretical limits of the completely saturated and unsaturated
absorptions ($1.0\le W_r\lambda1548/W_r\lambda1551 \le 2.0$).
\\
\\

\acknowledgements  We thank the anonymous referee for helpful
comments and suggestions. This work was supported by the National
Natural Science Foundation of China (NO. 11363001; No. 11073007),
the Guangxi Natural Science Foundation (2012jjAA10090), the
Guangzhou technological project (No. 11C62010685), and the Guangxi
university of science and technology research projects (NO.
2013LX155).

Funding for SDSS-III has been provided by the Alfred P. Sloan
Foundation, the Participating Institutions, the National Science
Foundation, and the U.S. Department of Energy Office of Science. The
SDSS-III web site is http://www.sdss3.org/.

SDSS-III is managed by the Astrophysical Research Consortium for the
Participating Institutions of the SDSS-III Collaboration including
the University of Arizona, the Brazilian Participation Group,
Brookhaven National Laboratory, Carnegie Mellon University,
University of Florida, the French Participation Group, the German
Participation Group, Harvard University, the Instituto de
Astrofisica de Canarias, the Michigan State/Notre Dame/JINA
Participation Group, Johns Hopkins University, Lawrence Berkeley
National Laboratory, Max Planck Institute for Astrophysics, Max
Planck Institute for Extraterrestrial Physics, New Mexico State
University, New York University, Ohio State University, Pennsylvania
State University, University of Portsmouth, Princeton University,
the Spanish Participation Group, University of Tokyo, University of
Utah, Vanderbilt University, University of Virginia, University of
Washington, and Yale University.

\end{document}